\documentclass[onecolumn]{emulateapj}
\lefthead{GAUDI \& WINN} \righthead{THE ROSSITER-MCLAUGHLIN EFFECT}
\def\eq#1{equation (\ref{#1})}
\def\Eq#1{Eq.~\ref{#1}}
\def\kms{{\rm km\,s^{-1}}}
\def\vpar{V}
\def\ms{{\rm m\, s^{-1}}}
\def\beq{\begin{equation}}
\def\eeq{\end{equation}}
\def\veq{V_{\rm eq}}
\def\ltsima{$\; \buildrel < \over \sim \;$}
\def\lsim{\lower.5ex\hbox{\ltsima}}
\def\gtsima{$\; \buildrel > \over \sim \;$}
\def\gsim{\lower.5ex\hbox{\gtsima}}

\begin{document}

\title{
Prospects for the Characterization and Confirmation of Transiting
Exoplanets via the Rossiter-McLaughlin Effect
}

\author {
B.\ Scott Gaudi\altaffilmark{1} and
Joshua N.\ Winn\altaffilmark{2}
}

\altaffiltext{1}{Harvard-Smithsonian Center for Astrophysics, 60
  Garden St., Cambridge, MA 02138}

\altaffiltext{2}{Department of Physics, Massachusetts Institute of
  Technology, 77 Massachusetts Ave., Cambridge, MA 02139}

\email{sgaudi@cfa.harvard.edu, jwinn@mit.edu}

\begin{abstract}

  The Rossiter-McLaughlin (RM) effect is the distortion of stellar
  spectral lines that occurs during eclipses or transits, due to
  stellar rotation. We assess the future prospects for using the RM
  effect to measure the alignment of planetary orbits with the spin
  axes of their parent stars, and to confirm exoplanetary transits. We
  compute the achievable accuracy for the parameters of interest, in
  general and for the 5 known cases of transiting exoplanets with
  bright host stars. We determine the requirements for detecting the
  effects of differential rotation. For transiting planets with small
  masses or long periods (as will be detected by forthcoming satellite
  missions), the velocity anomaly produced by the RM effect can be
  much larger than the orbital velocity of the star. For a terrestrial
  planet in the habitable zone of a Sun-like star found by the {\it
    Kepler} mission, it will be difficult to use the RM effect to
  confirm transits with current instruments, but it still may be
  easier than measuring the spectroscopic orbit.

\end{abstract}

\section{Introduction\label{sec:intro}}

When an exoplanet transits the disk of its parent star, there is both
a photometric signal and a spectroscopic signal. The photometric
signal is a small reduction in the received flux due to the partial
obscuration of the stellar disk, as first detected by Charbonneau et
al.\ (2000) and Henry et al.\ (2000). The transit light curve depends
chiefly on the radius of the planet, the radius of the star, the
orbital inclination, and the stellar limb darkening function. The
spectroscopic signal is subtler and less familiar. Changes in the
stellar absorption lines can be produced by absorption features in the
planetary atmosphere (see, e.g., Charbonneau et al.\ 2002,
Vidal-Madjar et al.\ 2003). Or, if the emergent stellar spectrum
varies with position across the stellar disk, then the planetary
obstruction will cause changes in the disk-integrated spectrum.

In particular, the Rossiter-McLaughlin (RM) effect refers to the
spectral distortion caused by the spatial variation in the emergent
spectrum due to stellar rotation. The exposed portion of the
photosphere has a net rotational Doppler shift. When the planet covers
part of the blueshifted half of the stellar disk, the integrated
starlight appears slightly redshifted, and vice versa. Thus, the
spectral distortion of the RM effect is often manifested as an
``anomalous'' radial velocity, i.e., a Doppler shift that is greater
or smaller than the shift expected from only the star's orbital
motion. Although the RM effect has a long history in the context of
eclipsing binary stars (Forbes 1911, Schlesinger 1911, Rossiter 1924,
McLaughlin 1924), its importance in the context of exoplanets is only
starting to be appreciated.

Suppose that the photometric signal has already been observed, and
that the parameters governing the photometric signal have been
accurately determined. The two most important additional parameters
that govern the RM effect are $V_S \sin I_S$, the projected rotation
speed of the stellar surface, and $\lambda$, the angle between the sky
projections of the stellar spin axis and the orbit normal (i.e., the
angle between the transit chord and lines of stellar latitude, for
$I_S=90\arcdeg$). Observations of the RM effect allow these two
parameters to be measured (see, e.g., Queloz et al.\ 2000, Ohta et
al.\ 2005, Winn et al.\ 2005, Wolf et al.~2006). The latter parameter,
$\lambda$, is especially interesting, as it provides information about
exoplanetary spin-orbit alignment. Solar system planetary orbits are
generally within $\sim$5\arcdeg\ of the solar equatorial plane (Beck
\& Giles 2005), but this may or may not be true for the full range of
exoplanetary systems.

Of particular interest in this regard are hot Jupiters---planets with
masses of $\sim$$M_{\rm Jup}$ and orbital periods of
$\sim$3~days---because there are some theoretical reasons why one
might expect large misalignments for such
planets.   Hot Jupiters are thought to have
formed at large orbital distances, and then migrated inward to their
current positions. The migration mechanism is still unknown, and some
of the proposed mechanisms differ in the degree to which they would
affect spin-orbit alignment. Thus, measurement of spin-orbit alignment
offers a possible means of discriminating among migration
theories. The most widely discussed category of migration theories
involves disk-planet interactions (Lin et al.\ 1996). There is a large
literature on this subject (as reviewed by Papaloizou \& Terquem
2006), but broadly speaking, there is no apparent reason why these
mechanisms would perturb spin-orbit alignment; in fact they may even
drive the system toward closer alignment. In contrast, other migration
theories involve disruptive events such as planet-planet interactions
(Rasio \& Ford 1996, Weidenschilling \& Marzari 1996) or planetesimal
collisions (Murray et al.\ 1998), which could act to randomize
spin-orbit alignment. Another proposal involves the Kozai mechanism,
in which a companion star causes oscillations in the planetary orbit's
eccentricity and inclination (Innanen et al.\ 1997, Holman et al.\
1997). By the time tides circularize the orbit and halt ``Kozai
migration,'' the orbital inclination can change substantially (Wu \&
Murray 2003; Eggenberger, Udry, \& Mayor 2004; D.\ Fabrycky \& S.\
Tremaine, priv.\ comm.).

For planets at larger orbital distances, there is no particular reason
to expect large misalignments, but the field of exoplanets has
rewarded observers with surprises in the past. There are also a few
empirical hints of multiple orbital planes in some systems, such as
the double debris disk recently reported around $\beta$~Pic
(Golimowski et al.\ 2006), and the apparently counter-rotating disks
around the protostar IRAS~16293--2422 (Remijan \& Hollis 2006).

The exoplanetary RM effect was first observed by Queloz et al.\ (2000)
and Bundy \& Marcy (2000) during transits of HD~209458b. The former
authors were able to place an upper bound on $\lambda$ of about
20\arcdeg. Snellen (2004) described the interesting idea of using the
wavelength-dependence of the RM effect to search for planetary
absorption features, but with only null results for HD~209458b. Winn
et al.\ (2005) used improved photometric and spectroscopic data to
show that $\lambda =-4^\circ\hskip-4.5pt.4\pm 1^\circ\hskip-4.5pt.4$,
a small but significant misalignment reminiscent of Solar system
planets.  \citet{wolf06} performed a similar study of HD~149026b,
finding $\lambda = 11\pm 14\arcdeg$. Apparently, in these two cases,
the migration mechanism was fairly quiescent. We can soon expect
similar studies of the other transiting planets with host stars bright
enough for high-precision spectroscopy, namely, HD~189733b
\citep{bouchy05c}, TrES-1 \citep{alonso04}, and XO-1b
\citep{mccullough06}. The first goal of this paper is to provide
useful guidance for future observations of these and other systems.

The second goal of this paper is to investigate another important and
timely application of the RM effect: transit detection and
confirmation. The spectroscopic detection of transits offers certain
advantages, in some cases, over photometric detection.  Indeed,
\citet{bouchy05c} discovered the transits of HD~189733b through
observations of the RM effect, and followed up this discovery with
photometric observations. The reason for this ordering may simply have
been that the observers had more convenient access to spectroscopic
resources than to photometric resources. However, as pointed out by
Welsh et al.\ (2004) and Ohta et al.\ (2005), the importance of the RM effect in transit
confirmation is likely to increase soon. The satellite missions {\it
Corot} \citep{baglin03} and {\it Kepler} \citep{borucki03} aim to find
smaller and longer-period transiting planets than those currently
known, for which it will be very difficult to confirm the existence of
transits photometrically from the ground. It has been envisioned that
the confirmation process will require measuring the orbital velocity
of the parent star, and there is hope that this can be achieved, given
the recent excellent progress in improving the accuracy of Doppler
measurements \citep{mayor03, marcy05}. As we will show in this paper,
the RM effect offers an alternative path to confirmation that is
suitable for at least a subset of stars.

Previous theoretical work on this topic has concentrated on analytic
descriptions of the spectral distortion (Ohta et al.\ 2005, Gimenez
2006), or on numerical simulations of the spectral distortion
involving a discretized model of the stellar surface (Queloz et al.\
2000, Welsh et al.\ 2004). In this work, we are not concerned with
high accuracy in describing the spectroscopic distortion; instead we
are concerned with the measurement and estimation problem. In
\S~\ref{sec:rmeffect}, we explain our notation, review some of the
previously derived results and provide some useful approximate scaling
relations. In \S~\ref{sec:character}, we estimate the achievable
accuracy in measuring the key RM parameters, $V_S \sin I_S$ and
$\lambda$, as a function of the orbital geometry of the system and of
the characteristics of the data. We do this both for the general case,
and for the specific cases of the 5 transiting exoplanets with bright
host stars. A secondary parameter that affects the RM signal is the
degree of differential rotation across the stellar disk, and in
\S~\ref{sec:limb} we investigate whether or not this effect is
important for near-term observations. In \S~\ref{sec:confirm}, we
derive an analytic expression for the signal-to-noise ratio ($S/N$) in
the detection of the RM effect, and we apply this formula to assess
the prospects for state-of-the-art Doppler measurements to confirm the
existence of transits of planets detected by {\it Kepler} and {\it
Corot}. The final section summarizes all of these results and suggests
some avenues for future work.

\section{The Rossiter-McLaughlin Effect}\label{sec:rmeffect}

\begin{figure}
\epsscale{1.2}
\plotone{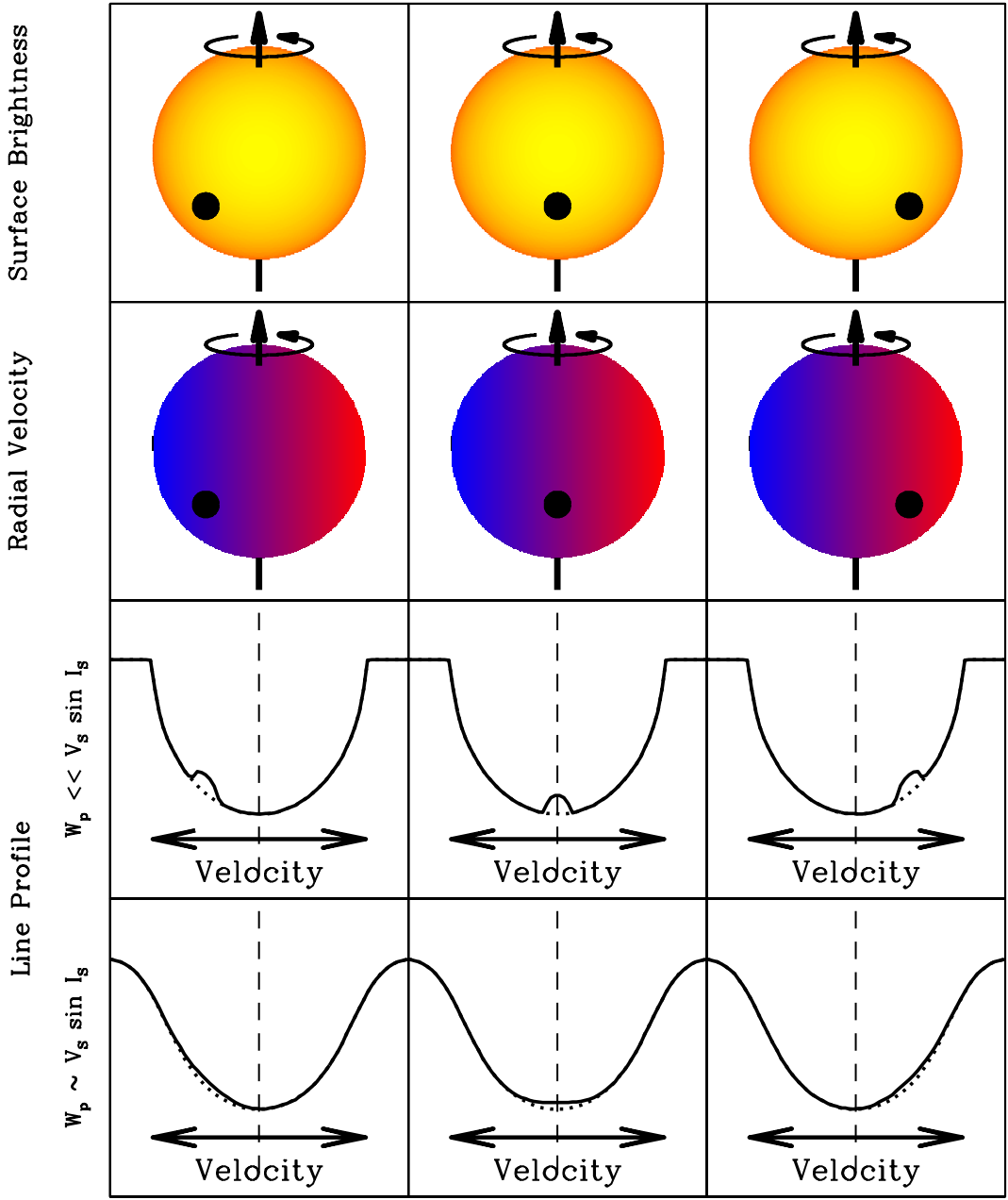}
\caption{
The physics of the RM effect.
{\bf Top row.} Three successive phases of an exoplanetary transit.
{\bf Second row.} Same, but the projected stellar rotation speed at each
point has been color coded, neglecting differential rotation.
At each phase, the planet hides a different velocity
component.
{\bf Third row.} Illustration of an observed stellar
absorption line, for the case of purely rotational broadening, i.e., the
net broadening due to all other mechanisms is much
less than the rotational broadening ($W_p \ll V_S \sin I_S$).
The missing velocity component
is manifested as a time-variable bump in the line profile.
{\bf Fourth row.} Same, but for the case
$W_p \sim V_S\sin I_S$, in which
other line-broadening mechanisms besides rotation are important.
\label{fig:speclines}
}\end{figure}

A pedagogic illustration of the physics of the Rossiter-McLaughlin
effect is presented in Figure~\ref{fig:speclines}. The top row of
panels illustrates the progress of a transiting planet across the
limb-darkened disk of its parent star.  This is the origin of the
photometric signal.  In the second row of panels, the stellar disk has
been color-coded according to the local line-of-sight velocity due to
rotation. One half is blueshifted, and the other half is
redshifted. This is the origin of the RM signal. The third row of
panels is a schematic view of a stellar absorption line that would be
recorded in unresolved (i.e., disk-integrated) observations, for the
case in which rotation is the dominant source of line broadening. The
planet hides a small fraction of a small range of velocity
components. The result is a ``bump'' that moves through spectral line
as the planet moves across the star. The idea is essentially the same
one that is employed in Doppler imaging of rapidly rotating stars
(see, e.g., Rice 2002), except in this case the contrast is provided
by an opaque foreground object rather than star spots. The final row
of panels depicts the case in which rotation is not the dominant
source of broadening. The distortion in this case is most easily
detected as a net Doppler shift.

Most of the known exoplanets have been discovered by searching for
periodic Doppler shifts in the spectrum of a star that result from the
star's reflex velocity (i.e., its orbital motion around the
star-planet center of mass). The target stars in these searches are
deliberately chosen to be inactive main-sequence FGK stars, which by
nature are slow rotators ($\lsim 5$~km~s$^{-1}$). The reason is that
those stars offer the smallest intrinsic velocity noise. This means
that the RM effect is appropriately described as an anomalous radial
velocity.  When a spectrum taken during transit is compared with a
standard stellar template in order to measure the radial velocity
shift, the distortion of the spectral lines will induce a signal which
appears as an anomalous radial velocity.

How can the size of the anomalous velocity be predicted, in terms of
the planetary and stellar properties? If, for example, the observed
spectrum is simply cross-correlated with a standard stellar template,
then there would be a wavelength shift in the peak of the
cross-correlation function relative to the unocculted case. In this
case, an excellent approximation for the velocity anomaly can be
obtained by computing the first moment of the distorted line profile
(i.e., the flux-weighted mean wavelength), and comparing it to the
first moment of the line profile in the out-of-transit spectrum, as
was done by Ohta et al.~(2005) and Gimenez~(2006).

In fact, as discussed by \citet{winn05b}, the accuracy of this
approximation depends on the exact procedure by which radial
velocities are extracted from a series of observed spectra. For
example, \citet{butler96} use a procedure that is optimized for high
precision velocity measurements using an iodine reference cell. It is
considerably more complicated than a simple cross-correlation, and has
as its working assumption that all spectral changes are due only to an
overall Doppler shift (which is to be measured) and variations in the
focus or the instrument. \citet{winn05b} gave evidence that in this
case the first-moment approximation for the RM effect is only accurate
to within $\sim$10\% for HD~209458. This level of accuracy is
sufficient for our main purpose of providing scaling relations and
estimated measurement accuracies within a factor of 2. For this
reason, we will employ the first-moment approximation throughout this
paper. However, it is interesting to note that the reported size of
the anomaly will depend on the precise method by which radial
velocities are extracted, which may make it difficult to compare the
results of different investigators even when they observe the same
system.

Consider a star with mass $M$ and radius $R$ that has a transiting
planet with mass $m$ and radius $r$. The orbit has a period $P$,
eccentricity $e$, and argument of pericenter $\omega$.  We can write
the net radial velocity variation of the star ($\Delta V$) as the sum
of the radial velocity of the star due to its orbital motion ($\Delta
V_O$) and the anomalous radial velocity due to the RM effect ($\Delta
V_R$): 
\beq
\Delta V(t) = \Delta V_O(t) + \Delta V_R(t).
\label{eqn:dv}
\eeq
The ``O''
stands for orbit, and the ``R'' refers to both Rossiter-McLaughlin and
rotation.

The line-of-sight component of the orbital velocity is
\beq
\Delta V_O(t) = K_O \left\{\cos[f(t)+\omega] + e \cos \omega\right\},
\label{eqn:dvd}
\eeq
where $f$ is the true anomaly and $K_O$ is the orbital velocity semiamplitude,
\beq
K_O= \left(\frac{2 \pi G}{P}\right)^{1/3} \frac{ m \sin I }{(M+m)^{2/3}}\left(1-e^2\right)^{-1/2}
\label{eqn:kd}
\eeq
$$
=8.9~{\rm cm\, s^{-1}} \left(\frac{P}{{\rm yr}}\right)^{-1/3}\left(\frac{m \sin i}{M_\oplus}\right)
\left(\frac{M}{M_\odot}\right)^{-2/3},
$$
where $I$ is the orbital inclination with respect to the sky plane. 
In the latter equality we have assumed $m\ll M$ and $e=0$.  

Assuming that the width of the absorption line is dominated by
rotational broadening, and further assuming that the stellar Doppler
shift is small, the first-moment approximation mentioned previously
gives \citep{ohta05}:
\beq
\Delta V_R(t) = -V_S \sin I_S \frac{\int\int x {\cal I}(x,y) dx dy}{\int\int {\cal I}(x,y) dx dy}.
\label{eqn:dvrm}
\eeq
Here, $V_S$ is the equatorial rotation speed of the stellar photosphere, $I_S$ is
the inclination of the stellar spin axis relative to the sky plane,
and ${\cal I}(x,y)$ is the surface brightness of the observed stellar
disk (including the dark spot due to the planet). The sky-plane
coordinates $x$ and $y$ are measured in units of the stellar radius,
have their origin at the projected center of the star, and are
perpendicular and parallel to the projected stellar rotation axis,
respectively. In fact, \eq{eqn:dvrm} also holds for lines that have
additional broadening mechanisms, such as thermal broadening, provided
that the additional broadening mechanisms produce no net Doppler shift
(i.e., the broadening kernel is symmetric about its centroid).

For convenience, we write the RM effect as
\beq
\Delta V_R(t) = K_R~g(t;x_p,y_p,\gamma,\epsilon,...),
\label{eqn:dvrm2}
\eeq
separating the overall amplitude $K_R$ of the RM effect from the
dimensionless function $g(t)~\la~1$. The amplitude is given by
\beq
K_R \equiv V_S \sin I_S~\frac{\gamma^2}{1-\gamma^2},
\label{eqn:krm}
\eeq
$$ = 52.8~\ms
\left(\frac{V_S\sin I_S}{5~\kms} \right)
\left(\frac{r}{R_{\rm Jup}}\right)^2 
\left(\frac{R}{R_\odot}\right)^{-2} 
$$
where $\gamma\equiv r/R$. In the latter equality, we have assumed
$\gamma \ll 1$.  For convenience, we will define $\vpar\equiv V_S\sin
I_S$. The dimensionless function $g$ depends primarily on the
projected position of the planet $(x_p,y_p)$, but also on $\gamma$ and
the limb-darkening function.  For simplicity, we use a
single-parameter ``linear'' description of the limb darkening law,
such that the (unocculted) surface brightness of the star is
\beq
\frac{{\cal I}(x,y)}{{\cal I}_0} =
1 - \epsilon
    \left[1-\left(1-x^2-y^2\right)^{1/2}\right],
\label{eqn:ld}
\eeq
with $\epsilon$ the linear limb darkening parameter. 
Note that in some circumstances---for example, the case of
differential rotation as discussed in \S\ref{sec:character}---the
function $g$ will depend on additional parameters.

\begin{figure}[t*]
\epsscale{1.0}
\plotone{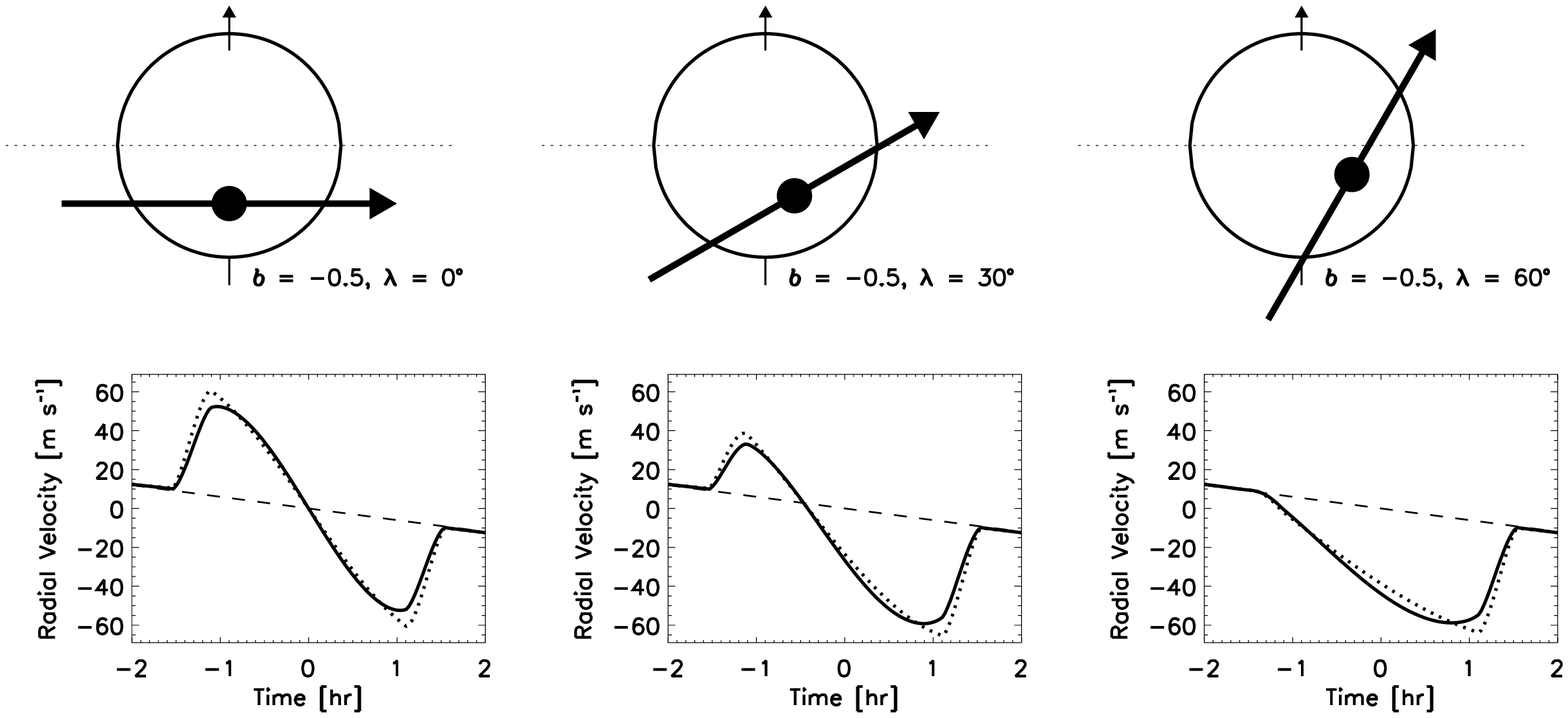}
\caption{
The dependence of the RM waveform on $\lambda$.
Three different possible trajectories of a transiting planet
are shown, along with the corresponding RM waveform (as computed
with the formulae of Ohta et al.~2005). The trajectories
all have the same impact parameter and produce the same
light curve, but they differ in $\lambda$ and produce
different RM curves. The dotted lines are for the case
of no limb darkening ($\epsilon=0$), and the solid lines are for
$\epsilon=0.6$.
\label{fig:trajectories}
}\end{figure}

Figure~\ref{fig:trajectories} shows three different trajectories of a
transiting planet across the stellar disk. These trajectories all have
the same impact parameter $b$, and consequently they all produce
exactly the same photometric signal.\footnote{The impact parameter is
  given by $b=a\cos I/R$, where $a$ is the orbital
semimajor axis.} However, the trajectories differ in the value
of $\lambda$, and consequently produce different RM waveforms, as
plotted in the lower row of panels. The sensitivity of the RM waveform
to $\lambda$ is what enables the observer to assess spin-orbit
alignment. The question of the achievable accuracy in $\lambda$ will
be taken up in the next section.

An especially simple case is when the planetary disk is fully
contained within the stellar disk, and limb darkening is negligible
($\epsilon=0$). In that case, $g$ is the perpendicular distance from
the projected stellar spin axis, $g(t)=x_p(t)$.  If we consider a
rectilinear trajectory across the face of the star with impact
parameter $b$, we can write the position of the center of the planet
as a function of time as,
\begin{eqnarray}
x_p(t) & = & \tau \cos \lambda - b \sin\lambda, \nonumber \\
y_p(t) & = & \tau \sin \lambda + b \cos\lambda,
\label{eqn:xpyp}
\end{eqnarray}
where $\tau\equiv (t-t_{\rm tra})/T_{\rm tra}$, $t_{\rm tra}$ is the time of
the transit midpoint, $T_{\rm tra}=R/v_{\rm orb}$ is the radius crossing time
corresponding to the planet's orbital velocity at the time of
transit (so that the transit duration is 
approximately $2T_{tra}\sqrt{1-b^2}$) and $\lambda$ is the angle of the trajectory with respect to
the the apparent stellar equator. We define $\lambda$ to be between
$-180\arcdeg$ and $+180\arcdeg$, such that for $\lambda>0$, the planet
moves towards the stellar north pole as it proceeds across the stellar
disk.  We note that due to the rotation of the star, the familiar symmetry
$b \leftrightarrow -b$ of the photometric signal is broken, and thus it is 
important to specify the sign of $b$ when considering the RM effect.

From \eq{eqn:xpyp}, it is clear that in the case of uniform rotation,
no limb-darkening, and when the planet is fully contained within the
stellar disk, the RM curve is simply a linear function of time. During
ingress and egress, and when limb darkening is taken into account, the
expression for $g$ is more complicated and may need to be evaluated
numerically. However, \citet{ohta05} and \citet{gimenez06} provide
useful approximate analytic expressions.

For our order-of-magnitude calculations, we will assume that the
planet is small ($\gamma \ll 1$), and we will not be concerned with
the ingress and egress phases because they constitute only a small
fraction ($\sim$2$\gamma$) of the entire duration of the transit. For
now, we will also neglect limb darkening. We therefore adopt the
simple approximation
\beq
g(t)=\left\{
\begin{array}{ll}
x_p(t)\qquad &{\rm if}\,\, \sqrt{\tau^2+b^2}\le 1,\\
0\qquad &{\rm otherwise}.
\end{array}\right.
\label{eqn:gt}
\end{equation}

It is interesting to compare the amplitude of the stellar orbital
velocity $K_O$ with the amplitude of the anomalous velocity $K_R$.
Assuming circular orbits ($e=0$), $\gamma \ll 1$, and $I =
90^\circ$, we find
\beq
\frac{K_R}{K_O} = \left(\frac{P \vpar^3}{2\pi G m}\right)^{1/3}
                  \left(\frac{\rho_*}{\rho_p}\right)^{2/3},
\label{eqn:kdkrm}
\eeq
where $\rho_p$ and $\rho_*$ are the average densities of the planet
and star, respectively. Since we expect $\rho_* / \rho_p$ to vary by
no more than a factor of a few from system to system, the order of
magnitude of $K_R/K_O$ depends mainly on the orbital period, the mass
of the planet, and the projected rotation speed of the star. We find
\beq
\frac{K_R}{K_O} \sim 0.3 
\left(\frac{m}{M_{\rm Jup}}\right)^{-1/3} 
\left(\frac{P}{3~{\rm days}}\right)^{1/3} 
\left(\frac{\vpar}{5~\kms}\right)
\label{eqn:kdkrmeval}
\eeq All of the currently-known transiting exoplanets have masses that
are comparable to Jupiter's mass, and orbital periods of 1--4~days.
For these systems, the anomalous velocity is smaller than the orbital
velocity by a factor of a few. However, the properties of the known
systems have been subject to very strong selection effects: the
transits of large, short-period planets are much easier to detect than
those of small, long-period planets (Gaudi~2005, Gaudi et
al.~2005). An interesting implication of \eq{eqn:kdkrmeval} is that
for the most challenging systems (small planets, long periods), the
amplitude of the RM effect will {\it exceed} the stellar orbital
velocity. In particular, for an Earth-mass planet with a period of one
year, $K_R/K_O \sim 3$ for $\vpar=5~\kms$. This explains the
appealing possibility of using the RM effect to detect or confirm
transits, which will be discussed further in \S~\ref{sec:confirm}.

\section{Prospects for Measuring Spin-Orbit Alignment}\label{sec:character}

In this section, we consider what can be learned from observations of
the RM effect in the regime of large $S/N$.  We have in mind
high-cadence, high-precision observations of both the photometric and
spectroscopic transit by a hot Jupiter, for which the spectroscopic
orbit is already well established.

As mentioned in \S~1, a principal goal of such studies is the
determination of $\lambda$, the angle on the sky between the stellar
spin axis and the planetary orbit normal, because this angle gives a
lower bound on any misalignment between the angular momenta of the
star and the orbit. The value of $\lambda$ is determined or bounded by
fitting a parameterized model to the photometric and spectroscopic
measurements (see, e.g., Winn et al.\ 2005 and Wolf et al.\
2006). However, for simple calculations, and for planning purposes, it
is useful to have some heuristics and order-of-magnitude estimates for
the effect of $\lambda$ on the RM waveform.

In particular, it is useful to consider the relative timing of three
observable events: the moment of greatest transit depth ($t_{\rm
  tra}$); the moment when the orbital radial velocity variation is
zero ($t_{\rm orb}$); and the moment when the anomalous RM velocity
variation is zero ($t_{\rm rot}$). At $t=t_{\rm tra}$, the projected
planet-star distance is smallest.  At $t=t_{\rm orb}$, the star is
moving in the plane of the sky.  At $t=t_{\rm rot}$, the planet lies
directly in front of the stellar rotation axis.  For a circular orbit
with $\lambda=0$, these three events are simultaneous. If the orbit is
circular but $\lambda \neq 0$, then $t_{\rm tra} = t_{\rm orb}$ but
$t_{\rm rot}$ will occur either earlier or later:
\begin{equation}
\frac{t_{\rm rot} - t_{\rm tra}}{T_{\rm tra}} \approx b \tan\lambda.
\label{eqn:deltat}
\end{equation}
This is easily derived from \eq{eqn:xpyp}.  For
a noncircular orbit, $t_{\rm tra}$ and $t_{\rm orb}$ are no longer
simultaneous.  To first order in $e$,
\begin{equation}
t_{\rm orb} - t_{\rm tra} = \frac{P}{2\pi} \left(e\cos\omega\right),
\label{eq:timing-offset}
\end{equation}
where $\omega$ is the argument of pericenter.  To the same order, both
the time difference $t_{\rm rot}-t_{\rm tra}$ and the transit duration
$T_{\rm tra}$ are multiplied by the same factor $1-e\sin\omega$, and
hence Eq.~\ref{eqn:deltat} remains valid.  These expressions give some
sense of the timing accuracy that is needed for a desired accuracy in
$\lambda$.  For example, for a mid-latitude transit at $b=0.5$ lasting
2.5~hr, a misalignment of $\lambda=1\arcdeg$ corresponds to a timing
offset of 45 seconds between the transit midpoint and the null in the
RM waveform.

\begin{figure}
\plotone{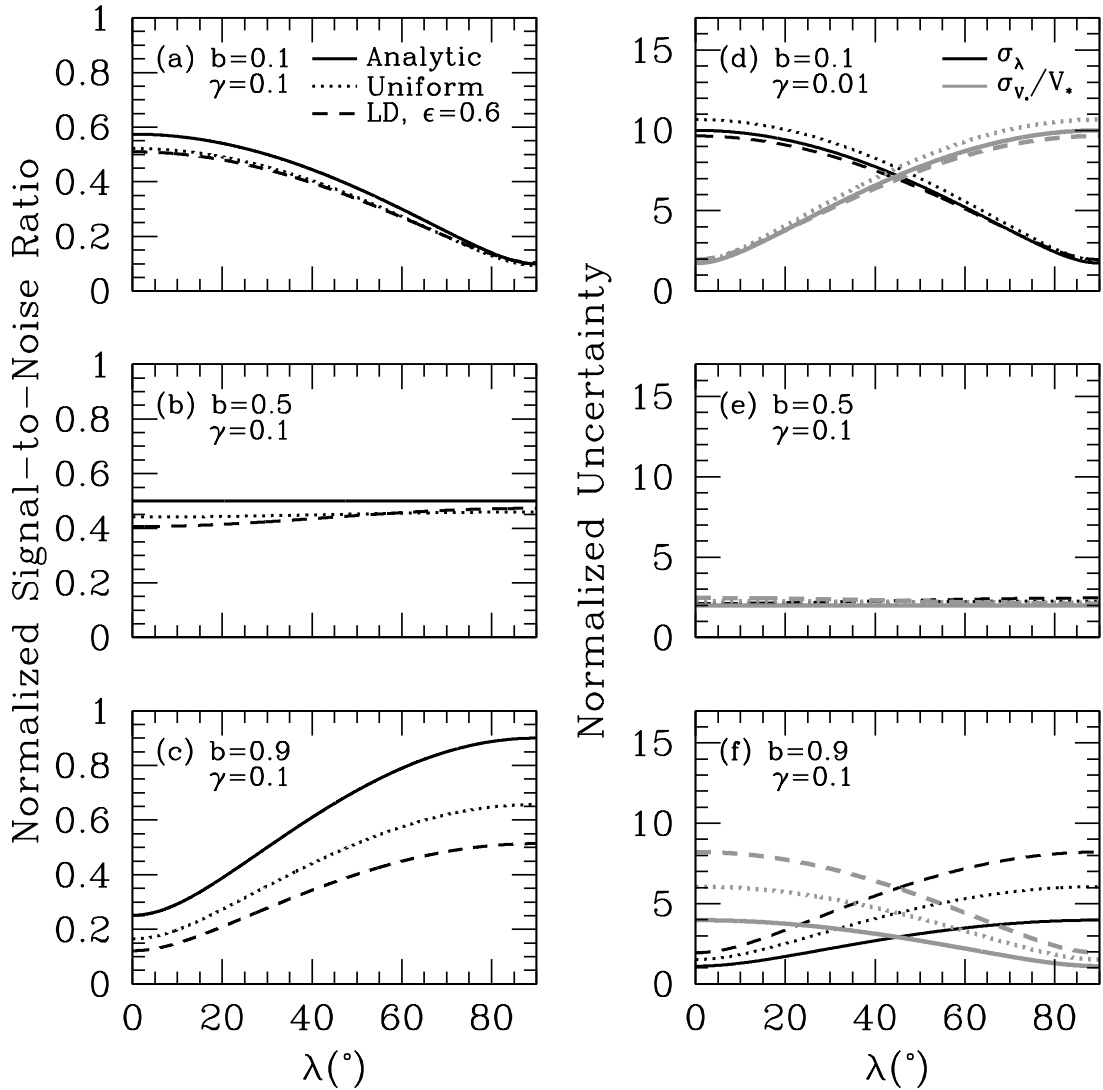}
\caption{
The achievable total signal-to-noise ratio in RM waveform (left), and
the signal-to-noise ratio in the the RM parameters $\lambda$ and
$\vpar$ (right), as a function of $b$ and $\lambda$.  Plotted is the
``normalized'' signal-to-noise ratio (i.e., after dividing by $Q_R$;
see Eq.~\ref{eqn:QRM}) for $\lambda$ (black lines) and for $V$ (gray
lines). The solid lines are based on our analytic formulae.  The
dotted and dashed lines are based on numerical computations with the
more accurate formulas of Ohta et al.~(2005); the dotted lines are for
the case of no limb darkening, and the dashed lines are for linear
limb darkening with $\epsilon=0.6$.
\label{fig:unc_lambda_v}
}
\end{figure}

Next, we derive an expression for the expected uncertainty in
$\lambda$ based on a series of spectroscopic measurements.  Consider a
series of $N$ radial velocity measurements taken at times $t_k$ during
the planetary transit (between first and fourth contact), each of
which has an uncertainty $\sigma$.  We consider the case in which the
RM measurements are the limiting source of error; we assume that both
the photometric transit and the spectroscopic orbit have already been
measured accurately.  Thus the times of contact, $t_{\rm tra}$, and
the parameters $\gamma$ and $b$ have negligible uncertainties, and the
orbital velocity $\Delta V_O(t)$ can be accurately subtracted from the
total velocity variation observed during transits to isolate the RM
waveform.  The only parameters to be determined by fitting the transit
data are $\lambda$ and $\vpar$, which we combine into a
two-dimensional parameter vector $\vec{a}$. The expected uncertainties
in $\lambda$ and $\vpar$ are the square roots of the diagonal elements
of a matrix $C$ that is given by
\begin{equation}
C=B^{-1},
\label{eqn:cmat}
\end{equation}
where $B$ is related to the Fisher information matrix (see
\citealt{gould03}) and is calculated as
\begin{equation}
B_{ij} \equiv \sum_{k=1}^N \frac{1}{\sigma^2} 
\left[\frac{\partial}{\partial a_i}  \Delta V_R(t_k) \right]
\left[\frac{\partial}{\partial a_j}  \Delta V_R(t_k) \right] .
\label{eqn:fisher}
\end{equation}
We will adopt the approximate analytic form for $\Delta V_R(t)$ that
was given in \eq{eqn:gt}.  Assuming evenly-spaced observations and
large $N$, we can convert the sum in \eq{eqn:fisher} into an integral.
This then yields the following expressions for the expected
uncertainties in $\lambda$ and $\vpar$:
\begin{equation}
\sigma_\lambda = Q_R^{-1}
\left[\frac{(1-b^2)\sin^2{\lambda}+3 b^2 \cos^2{\lambda}}{b^2(1-b^2)}\right]^{1/2},
\label{eqn:sigl}
\end{equation}
\begin{equation}
\frac{\sigma_{\vpar}}{\vpar} = Q_R^{-1}
\left[\frac{(1-b^2)\cos^2{\lambda}+3 b^2 \sin^2{\lambda}}{b^2(1-b^2)}\right]^{1/2},
\label{eqn:sigvpar}
\end{equation}
where we have defined
\begin{equation}
Q_R \equiv \sqrt{N}\frac{K_R}{\sigma}.
\label{eqn:QRM}
\end{equation}
As discussed further in \S\ref{sec:confirm}, the factor $Q_R$ is
proportional to the total signal-to-noise ratio of the measured RM
waveform.  Note that $\sigma_{\vpar}(\lambda)/\vpar
=\sigma_\lambda(\pi-\lambda)$.
Similarly, we can derive the covariance between $\lambda$ and
$\vpar$, which is given by ${\rm
  cov}(\lambda,\vpar)=C_{12}/(C_{11}C_{22})^{1/2}$.  We find
\beq
{\rm cov}(\lambda,\vpar)=
\frac{4b^2-1}{\left[(1-b^2)^2+9b^4+3b^2(1-b^2)(\tan^2\lambda+\cot^2\lambda)\right]^{1/2}}.
\label{eqn:covlv}
\eeq The uncertainty in each parameter is the product of $Q_R^{-1}$
and a factor that depends on the orbital geometry. The geometrical
factor is illustrated in Figure \ref{fig:unc_lambda_v}, for
$\gamma=0.1$ and three values of the impact parameter $b$. The solid
lines show the expected uncertainties in $\lambda$ and $\vpar$ given
by equations (\ref{eqn:sigl}) and (\ref{eqn:sigvpar}), after dividing
by $Q_R^{-1}$.  The three cases are a near-central transit ($b=0.1$),
a mid-level transit ($b=0.5$), and a grazing transit ($b=0.9$). For
$b=0.5$, we find that $\sigma_\lambda=\sigma_{\vpar}/\vpar=2Q_R^{-1}$,
and that the uncertainties in $\vpar$ and $\lambda$ are uncorrelated
[${\rm cov}(\lambda,\vpar) = 0$].  In this sense, mid-level transits
are ideal for cleanly separating the effects of $V_S\sin I_S$ and
$\lambda$ on the RM waveform.

In contrast, for nearly central transits there is a strong degeneracy
between $\lambda$ and $\vpar$ and a strong dependence of
$\sigma_\lambda$ on $\lambda$.  For $b=0$ exactly, ${\rm
  cov}(\lambda,\vpar)=-1$. The observed signal depends only on the
parameter combination $\vpar\cos{\lambda}$, and the timing offset
given in Eq.~(\ref{eq:timing-offset}) is zero regardless of $\lambda$.
One can measure only the amplitude of the RM waveform, and cannot tell
whether a small amplitude (say) is the result of an equatorial transit
across a slowly rotating star, or a misaligned transit across a more
rapid rotator.  An accurate measurement and interpretation of the line
broadening in the out-of-transit spectra is essential here, by
providing an independent estimate of $\vpar$ that can be used as an
{\it a priori} constraint. For grazing transits, there is a more
modest dependence of the uncertainties on $\lambda$ (although our
approximations are least accurate for grazing transits, as described
below).

In order to verify our analytic expressions, and evaluate the
importance of some of the effects that these expressions neglect
(namely, the finite ingress/egress durations and limb darkening), we
numerically evaluated the elements of the Fisher matrix $B$ using the
more accurate but more complex expressions for $\Delta V_R(t)$ given
by \citet{ohta05}. The results are also plotted in Figure
\ref{fig:unc_lambda_v}.  The dotted lines show the results for
$\epsilon=0$ (no limb darkening), and the dashed lines show the
results for $\epsilon=0.6$.  Except for non-grazing transits, the
differences between the uncertainties predicted by our analytic
expressions and the numerically-calculated uncertainties are small
($\la 10\%$).  Not surprisingly, for grazing transits the differences
are substantially larger and can be a factor of 2.

Next, we verified our results using Monte Carlo simulations.  For a
given choice of $\vpar$ and $\lambda$, we created $5000$ simulated
data sets of radial velocity measurements during a transit.  Each data
point was the value calculated according to the expressions of Ohta et
al.~(2005), plus Gaussian noise with a standard deviation of
$\sigma$.  We then used a downhill-simplex algorithm to optimize the
parameter values $\vpar$ and $\lambda$ for each data set.  We
calculated the dispersions in the resulting distribution of 5000
fitted values, and took the dispersion to be the uncertainty in that
parameter. We repeated this procedure for a range of $\vpar$ and
$\lambda$.  The resulting uncertainties agree very well with the
uncertainties computed by numerically evaluating the elements of $B$
as described above.

We now consider the 5 particular cases of known transiting
exoplanets with bright ($V<12$) parent stars: HD~209458 (Henry et
al.~2000, Charbonneau et al.~2000), TrES-1 (Alonso et al.~2004),
HD~189733 \citep{bouchy05c}, HD~149026 (Sato et al.~2005), and
XO-1 (McCullough et al.~2006). For the systems for which the RM effect
has already been measured, this exercise provides an empirical check
on our calculations. For the others, it provides a guide for the
achievable accuracy of future observations.  Below we describe our 
assumptions and the results for each planet in detail; this information
is summarized in Table \ref{tab:predictions}. 

For HD~209458, an analysis of the best available photometric and
spectroscopic data was performed by \citet{winn05b}.  The data set
included $N=14$ radial velocity measurements during transits, with a
typical uncertainty of $\sigma=4.1~\ms$.  By fitting a parameterized
model to all of the data, and evaluating the parameter uncertainties
with a Monte Carlo bootstrap technique, they found $\lambda
=-4^\circ\hskip-4.5pt.4\pm 1^\circ\hskip-4.5pt.4$ and $\vpar= 4.70 \pm
0.16~\kms$.  Using these values, as well as the parameters
$\gamma=0.121$ and $b=0.52$ determined from the photometry, our
analytic estimates predict that the uncertainties should be
$\sigma_\lambda=1^\circ\hskip-4.5pt.7$ and
$\sigma_\vpar=0.15~\kms$. These estimates are in excellent agreement
with the actual uncertainties derived from detailed model-fitting.  In
addition, Eq.~(\ref{eqn:covlv}) predicts that the uncertainties in
$\vpar$ and $\lambda$ should be uncorrelated, as was indeed found to
the be the case by Queloz et al.~(2000) and Winn et al.~(2005).

For HD~149026, \citet{wolf06} analyzed all of the available
photometric and spectroscopic data, including $N=15$ radial velocity
measurements taken during transit. The typical velocity uncertainty
was $\sigma=4.0~\ms$. Through parameterized model-fitting and
bootstrap resampling, these authors found $\lambda
=11^\circ\hskip-4.5pt \pm 14^\circ\hskip-4.5pt$, $\vpar=
6.4_{-0.7}^{+2.1}~\kms$, $\gamma=0.051$, and $b=0.39$. Our analytic
formulae predict $\sigma_\lambda=9.0^\circ\hskip-3.5pt$ and
$\sigma_\vpar=0.76~\kms$. Our estimates are smaller than the
\citet{wolf06} uncertainties. We believe that the reason is that the
photometric signal is not nearly as well determined for HD~149026 as
it is for HD~209458, not only because of the smaller size of the
planet, but also because the existing observations are from
ground-based telescopes as opposed to the higher-precision {\it Hubble
Space Telescope} light curve obtained by Brown et al.~(2001) for
HD~209458.  Consequently, the uncertainties in the parameters $t_{\rm
tra}, \gamma$, and $b$ cannot be neglected.  The covariance with these
photometric parameters contribute to the uncertainty in $\lambda$ and
$\vpar$, violating one of the conditions for the accuracy of our
analytic estimates. We conclude that there is scope for improvement in
the determination of $\lambda$ for this system through improved
photometry.

As mentioned in the introduction, the transits of HD~189733b were
originally discovered via the RM effect \citep{bouchy05c}, but those
authors did not attempt to measure $\lambda$ or $\vpar$ by fitting the
RM waveform.  However, using their estimate of $\vpar=3.5\pm 1.0~\kms$
derived from the out-of-transit spectral line profile, we can
anticipate the expected uncertainties of such attempts, through an
application of our expressions.  The photon noise in the radial
velocity measurements reported by \citet{bouchy05c} amounts to only
$\sim$5-7~$\ms$, but the actual uncertainties are larger because the
star is fairly active.  As an estimate of this ``stellar jitter,'' we
will adopt the value $\sigma=15~\ms$ based on the scatter around the
Keplerian orbital solution. Taking $b=0.66$ and $\gamma=0.157$
\citep{bakos06}, we predict $\sigma_\lambda=5^\circ\hskip-4.5pt.2$ and
$\sigma_\vpar=0.48~\kms$ for $N=8$ (again, assuming that the
measurement of the RM waveform is the limiting uncertainty).  These
values are based on the assumption $\lambda\sim 0$, although the
expected uncertainties are fairly insensitive to $\lambda$ because the
transit is neither central nor grazing (see the previous section).  If
the velocity jitter turns out to be smaller, or if a larger number of
transit velocities are measured, then the uncertainties in the RM
parameters will scale as $\sigma/\sqrt{N}$.

No measurements of the RM effect have yet been reported for TrES-1 or
XO-1. We estimate the accuracy with which $\lambda$ can be constrained
in these systems with future observations, using the parameters of the
systems that have been derived from the existing photometric and
spectroscopic data.  For specificity, we assume $N=15$ and
$\sigma=15~\ms$.  The latter estimate for $\sigma$ seems reasonable,
given that the apparent magnitudes of the host stars ($m_V=11-12$) are
fainter than the three stars considered above ($m_V=8$).

For TrES-1, there exists only an upper limit on the projected stellar
rotation velocity of $\vpar\la 5~\kms$. We will optimistically assume
that the true value is $5~\kms$, right at the upper limit. Then,
assuming $b=0.18$ and $\gamma=0.137$ \citep{alonso04,sozzetti04}, we
find $\sigma_\lambda=13^\circ$ and
$\sigma_\vpar=0.36~\kms$ for $\lambda=0$, and
$\sigma_\lambda=4^\circ\hskip-4.5pt.1$ and $\sigma_\vpar=~1.1~\kms$ for
$\lambda=90^\circ$. The strong sensitivity to $\lambda$ arises because
the transit is nearly central.

For XO-1, we adopt $\vpar=1.11~\kms$, $b=0.12$ and $\gamma=0.131$
\citep{mccullough06,holman06}, and find $\sigma_\lambda=95^\circ$ and
$\sigma_\vpar=0.39~\kms$ for $\lambda=0$, and $\sigma_\lambda=20^\circ$
and $\sigma_\vpar=1.8~\kms$ for $\lambda=90^\circ$. Again, for this
fairly central transit, the expected uncertainty depends strongly on
$\lambda$.

For both TrES-1 and XO-1b the uncertainties in $\lambda$ and $\vpar$
are expected to be fairly large unless a substantial commitment of
resources are expended to acquire many high-$S/N$ spectra during
transit, or the individual velocity uncertainties can be reduced
substantially below the $15~\ms$ assumed here.  This is because in
both cases, the projected stellar rotation velocities are fairly
small, and the transits are nearly central (small $b$). This gives
rise to large uncertainties in $\vpar$ (for $\lambda \sim 0$) or
$\lambda$ (for $\lambda \sim 90^\circ$).

\begin{deluxetable}{cccccccccc}
\setlength{\tabcolsep}{0.1in}
\tablecaption{Predictions for Rossiter-McLaughlin Effect Parameter Uncertainties}
\tablewidth{0pt}
\tablehead{
  \colhead{Name} &
  \colhead{$N$} &
  \colhead{$\sigma$} &
  \colhead{$\gamma$} &
  \colhead{$b$} &
  \colhead{$\vpar$\tablenotemark{a}} &
  \colhead{$\lambda$} &
\colhead{$K_{R}$}&
 \colhead{$\sigma_\lambda$} &
  \colhead{$\sigma_\vpar$} \\
  \colhead{} &
  \colhead{} &
  \colhead{$\ms$} &
  \colhead{} &
  \colhead{} &
  \colhead{$\kms$} &
  \colhead{deg} &
   \colhead{$\ms$} &
 \colhead{deg} &
  \colhead{$\kms$} \\
}
\startdata
 HD 209456b & 14 & 4.1 & 0.121 & 0.52  & 4.70 & $-4.4$ & 70 & 1.7  & 0.15 \\
 HD 149026b & 15 & 4.0 & 0.051 & 0.39  & 6.4  & 11 & 17 & 9.0 & 0.76 \\
 HD 189733b &  8 & 15  & 0.157 & 0.66  & 3.5  &  0\  & 88 & 5.2 & 0.48 \\
   --       & -- & --  & --    & --    & --   &  90 & --  & 7.9 & 0.32 \\
 TrES-1b    & 15 & 15  & 0.137 & 0.18 & 5\tablenotemark{b} & 0 & 96 & 13 & 0.36 \\
 --         & -- & --  & --    & --    & --   &  90 & -- & 4.1 & 1.1 \\
 XO-1b      & 15 & 15  & 0.131 & 0.12  & 1.11 & 0 & 19 & 95 & 0.39 \\
   --       & -- & --  & --    & --    & --   &  90 & -- & 20 & 1.8 \\

\tableline
\enddata
\tablenotetext{a}{$\vpar\equiv V_S\sin I_S$}
\tablenotetext{b}{For TrES-1b, we assume a projected
stellar velocity that is equal to the spectroscopically determined upper
limit of $\vpar=5~\kms$.}
\tablecomments{For HD 189733b, a fit to the Rossiter-McLaughlin effect data
has not been reported, and so no constraint on $\lambda$ is available.  We therefore
show the expected parameter uncertainties for the two values $\lambda=0,90^\circ$, which will
bracket the range of uncertainties.  For TrES-1b and XO-1b, no Rossiter-McLaughlin
data has been reported, and so we assume $N=15$, $\sigma=15~\ms$, and 
show the expected parameter uncertainties for the two values $\lambda=0,90^\circ$.
}\label{tab:predictions}
\end{deluxetable}

In addition to these 5 systems, there are 5 other known transiting
planets, all of which were originally identified as candidates by the
OGLE collaboration \citep{udalski02a,udalski02b,udalski02c,udalski03},
and subsequently confirmed with radial velocity follow-up
\citep{konacki03a,konacki03b,konacki04,konacki05,
  bouchy04,pont04,moutou04,pont05,bouchy05a}. The stars in these
systems are all fainter than the five cases considered above
($m_V\simeq 15.5-17$ versus $m_V \simeq 8-12$), which will make the
measurement of the RM effect much more challenging for these
systems. Nevertheless, there is a good reason to expend this
additional effort on at least some of the OGLE systems, namely, the
``very hot Jupiters'' (VHJs) with orbital periods less than 3 days. It
has been suggested that these shorter-period planets discovered by
OGLE form a distinct population from the more numerous ``hot
Jupiters'' with 3-4~day periods, based on differences in their typical
mass \citep{gaudi05,mzp05} and their overall frequency
\citep{gaudi05,gould06}. This raises the interesting possibility that
the VHJs arrived at their very close orbits through a different
mechanism than the hot Jupiters. For example, the observation that the
orbital distances of VHJs are nearly equal to twice their Roche radii
\citep{ford06} may indicate that these planets were emplaced via tidal
capture and subsequent circularization. Such a scenario would be the
natural result of planet-planet scattering \citep{rasio96,ford01},
capture of free-floating planets \citep{gaudi03}, or Kozai
oscillations (\citealt{wu03}, D.\ Fabrycky \& S.\ Tremaine, priv.\
comm.). In the latter two scenarios, at least, one would expect these
planets to generically have orbits with large misalignments with the
spin axis of their parent stars.

We can repeat our previous analysis to estimate the achievable
uncertainty in $\lambda$ for the OGLE systems. Unfortunately, many of
the parameters needed to provide an accurate estimate (particularly
the impact parameter $b$ and projected stellar rotation speed $\vpar$)
have not been measured, or have substantial uncertainties. We will
therefore adopt approximate, fiducial values in order to provide an
order-of-magnitude estimate, with the caution that the actual
uncertainties could be significantly different.  We will assume
$b=0.5$, $\gamma=0.1$ and $\epsilon=0.6$.  Only upper limits for the
rotational velocities of the host stars have been reported, so we will
adopt the largest allowed velocity of $\vpar=5~\kms$.  Typical radial
velocity precisions reported for these systems are $\sigma \simeq
50~\ms$ for $\sim 45~{\rm minute}$ exposures (using 10m-class
telescopes), which we scale to $\simeq 100~\ms$ for $10$-minute
exposures.  Assuming continuous observations during a typical transit
duration of $\sim$2~hr (i.e., twelve $10$-minute exposures), we find
$Q_R \simeq 1.7$ per transit, and so $\sigma_\lambda = 2Q_R^{-1} \sim
70^\circ$.  With observations of 4 transits, one could measure
$\lambda$ to within about $35^\circ$.  While this is not nearly as
good as the results that can be achieved for the systems with brighter
host stars, even the finding that one of the VHJs has (say) $\lambda =
-90^\circ \pm 35^\circ$ would be quite interesting.

\section{Differential Rotation}\label{sec:limb}

In the preceding calculations we neglected differential rotation,
i.e., we assumed that the rotation speed of the star was independent
of the stellar latitude. One might wonder whether or not the expected
degree of differential rotation is detectable through transit
observations. If so, it would be important both as a stellar
astrophysics tool, and (if it is not modeled properly) as a possible
source of ambiguity or covariance in the interpretation of RM data.

Differential rotation is an important phenomenon in stellar
astrophysics because it is an observable manifestation of convective
dynamics, and because it is intimately linked to the dynamo mechanism
of stellar magnetic field generation and evolution.  The Sun rotates
$\sim$20\% faster at its equator than at high latitudes (see, e.g.,
Howard 1984 and references therein). Differential rotation has also
been measured for other stars using star spots. In some cases, spots
at different latitudes produce measurably different photometric
periods (see, e.g., Henry et al.~1995, Rucinski et al.~2004, Herbst et
al.~2006), and in other cases, the motion of star spots has been
tracked via Doppler imaging (Collier Cameron et al.~2002).

In principle, transiting exoplanets can be used instead of star spots.
The advantages would be that planetary orbits are much more stable
than star spot patterns, the transit chord can occur at any stellar
latitude and can span a wide range of latitudes (unlike star spots),
and the time scale of the transit (a few hours every few days) is more
convenient than the month-long rotation periods of many stars. In
addition, it would allow differential rotation to be measured on
inactive and unspotted stars for the first time.  A system of multiple
transiting planets with different impact parameters has yet to be
discovered, but would be a wonderful probe of differential
rotation. But even with a single transiting planet, it is possible in
principle to detect differential rotation through the RM effect if
$\lambda \neq 0$. The RM waveform effectively traces the projected
surface speed of the star along a one-dimensional chord, and is
sensitive to differential rotation if the chord spans a significant
range of latitude.

It is customary to parameterize observations of differential rotation
according to a function such as $V(\ell)=\veq(1 - \alpha
\sin^2{\ell})$, where $\veq$ is the rotation speed at the stellar
equator, $\ell$ is the latitude on the surface of the star, and
$\alpha$ is the differential rotation parameter (which is equal to the
fractional difference in rotation speed between the pole and the
equator). For the Sun, differential rotation is significant,
$\alpha=0.2$. For stars with convective envelopes ($T_{eff} \la
7500~{\rm K}$), the amount of differential rotation is thought to be
correlated with $\veq$, such that more rapid rotators exhibit less
differential rotation (Barnes~2005, Reiners~2006).  For stars lacking
exterior convection zones, the surface differential rotation is small
and $\alpha \sim 0$.

If the surface of the star is in uniform rotation, then the surface
speed at any point on the star depends only on the projected rotation
speed of the star, $\vpar=V_S \sin I_S$, and the distance from the the
projected rotation axis, $x_p$ (\Eq{eqn:xpyp}).  However, in the case
of differential rotation, this symmetry is broken. The surface speed
at a given point on the star also depends on the position parallel to
the projected axis $y_p$, as well as on the inclination of the
rotation axis, $I_S$.  The 
apparent rotation speed at any point $(x_p,y_p)$ of the
stellar surface is
\begin{equation}
v(x_p,y_p) =
 -V_{\rm eq} \sin I_S~~x_p
 \left\{1-\alpha\left[y_p \sin{I_S}+(1-x_p^2-y_p^2)^{1/2}\cos{I_S}\right]^2\right\}.
\label{eqn:vxpyp}
\end{equation}

\begin{figure}[t*]
\epsscale{1.0}
\plotone{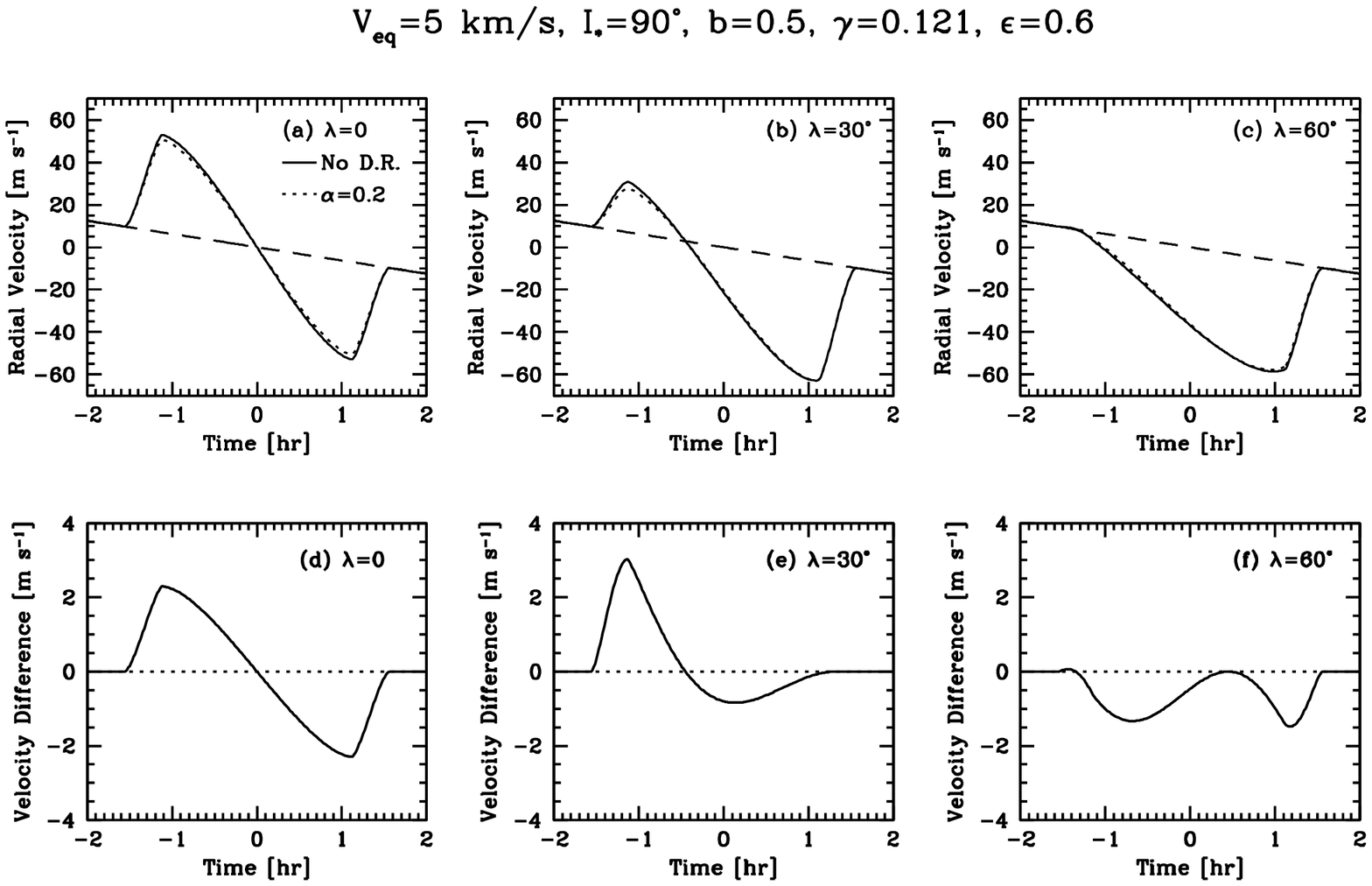}
\caption{
The spectroscopic transit in the presence of differential rotation.
The top row of panels shows the expected RM waveform for a transit
in a system similar to HD~209458 ($b=0.5$, $\gamma=0.121$, $\veq=5~\kms$),
for three different values
of $\lambda$. The stellar rotation axis is taken to be in the plane
of the sky, and the differential rotation parameter was set to $\alpha=0.2.$
The dotted lines show the corresponding waveforms for $\alpha=0$.
The bottom row of panels show the difference between the $\alpha=0.2$
and $\alpha=0$ cases.
\label{fig:diffrot}
}\end{figure}

Figure \ref{fig:diffrot} shows the effect of differential rotation on
the RM waveform. For this illustration, we used the parameters
appropriate for HD~209458 ($\gamma=0.121$, $b=0.5$, and $\veq=5~\kms$)
and further assumed $I_S=90^\circ$, so that the rotation axis is in
the plane of the sky. We took the differential rotation parameter to
be equal to the solar value of $\alpha=0.2$. Each column shows the
results for a different value of $\lambda$. The top row shows the
waveform itself, and the bottom row shows the difference between the
waveforms with $\alpha=0$ and $\alpha=0.2$.

For $\lambda=0$ and $I_S=90^\circ$, the effect of differential
rotation is degenerate with changing the value of $V_{\rm eq}$, and
hence differential rotation is not measurable.  However, from Figure
\ref{fig:diffrot} and \eq{eqn:vxpyp}, it is clear that for $\lambda\ne
0$ or $I_S \ne 90^\circ$, these two parameters are not degenerate, and
in principle both parameters can be constrained by fitting a
parameterized model to the velocity data. However, the requirement on
the velocity precision is stringent. The amplitude of the deviations
due to differential rotation generally scale with $\alpha$, and are
only a few m~s$^{-1}$ for $\alpha =0.2$.

\begin{figure}
\plotone{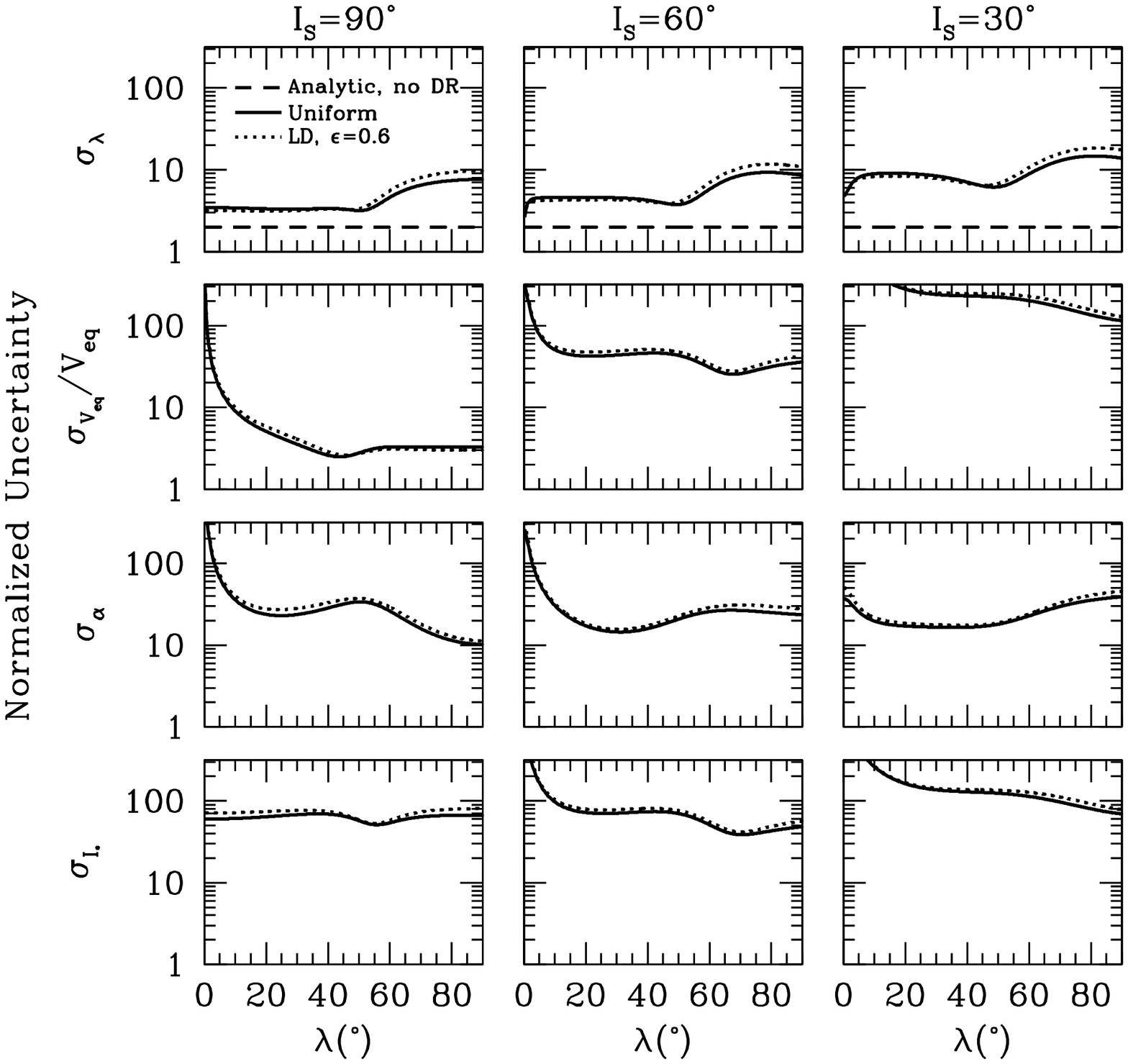}
\caption{ 
The achievable signal-to-noise ratio in the RM and differential
rotation parameters $\lambda, V_{eq}, \alpha, I_S$,
as a function of $\lambda$ for three values of $I_S$.  We have
assumed $b=0.5$, $\gamma=0.121$, and $\epsilon=0.6$, thus
approximating the HD~209458 system.  We have assumed
differential rotation parameter equal to the solar value of $\alpha=0.2$.  
Plotted is the ``normalized'' signal-to-noise ratio (i.e., after dividing by $Q_R$;
see Eq.~\ref{eqn:QRM}). The solid lines show
the case for no limb darkening, and the dashed lines are for linear
limb darkening with $\epsilon=0.6$. The dashed lines in the top row show
our analytic formula for the uncertainty in $\lambda$ in the
absence of differential rotation.  The difference between the dashed and 
solid/dotted lines demonstrates the degradation of the uncertainty
in $\lambda$ due to covariance with the differential rotation
parameters. 
\label{fig:drerrs}
}\end{figure}

For a more quantitative analysis, we can apply the same procedure that
we used in the previous section to estimate how well one can hope to
estimate $\alpha$ using the RM effect. We estimate the uncertainties
by numerically computing the Fisher matrix (\Eq{eqn:fisher}) using the
analytic form for the RM effect given by \citet{ohta05}, but
accounting for the effect of differential rotation on projected
surface velocity (\Eq{eqn:vxpyp}). We consider the uncertainties on
four parameters $\vec{a}=(\lambda, v_{\rm eq}, \alpha, I_S)$, and we
assume that all other parameters have negligible uncertainties.
The resulting uncertainties in the parameters $\vec{a}$ exhibit
complicated relationships with $\lambda$, $b$, and $I_S$.  We will
summarize our results by focusing on the case $\alpha=0.2$, $b=0.5$,
$\gamma=0.121$, and $\epsilon=0.6$ (approximating the HD~209458
system).  Figure \ref{fig:drerrs} shows the expected uncertainties as
a function of $\lambda$ for $I_S=30^\circ, 60^\circ, 90^\circ$.  We
show the ``normalized uncertainties,'' after dividing by $Q_R$, where
$Q_R$ is still given by \eq{eqn:QRM}, but with a slightly altered
definition of RM semiamplitude from \eq{eqn:krm}: $K_R \equiv \veq
{\gamma^2}/({1-\gamma^2})$.

First, we consider the case in which the stellar rotation axis is in
the plane of the sky: $I_S=90^\circ$ (left column of Fig.\
\ref{fig:drerrs}).  As seen in Figure \ref{fig:diffrot}, for this case
and $\lambda=0$, $\alpha$ is degenerate with $\veq$, and it is not
possible to determine these parameters independently. In fact, for
$I_S=90^\circ$ and $\lambda=0$, neither of the differential rotation
parameters ($\alpha, I_S$) are well constrained.  We find that in
general the uncertainty in $\lambda$ is degraded by covariances with
the other parameters (particularly $\alpha$).  For example, we find
$\sigma_\lambda \sim 3 Q_R^{-1}$ for $\lambda \sim 0$.  This is $\sim
50\%$ larger than the uncertainty expected under the assumption of
solid body rotation (\Eq{eqn:sigl}). Thus without an {\it a priori}
constraint on the allowed degree of differential rotation,
measurements of $\lambda$ will be significantly compromised by an
unknown amount of differential rotation.  We explore the effects of a
constraint on $\alpha$ in more detail below.

From inspection of Figure \ref{fig:drerrs}, we see that for
$I_S=90^\circ$, the most favorable value of $\lambda$ for measuring
the overall system parameters (in the sense of minimizing the sum of
the squares of the uncertainties of all four parameters) is
$\lambda\approx 55^\circ$. For this value, we find
$\sigma_\lambda=4.1Q_R^{-1}$, $\sigma_{V_{eq}}/V_{eq}=3.0Q_R^{-1}$,
$\sigma_\alpha=35Q_R^{-1}$, and $\sigma_{I_S}=52Q_R^{-1}$.  Therefore
measuring $\alpha$ to within 0.2 and $I_S$ to within $15^\circ$
requires $Q_R\sim 200$.  Achieving this high a signal-to-noise ratio
for a system similar to HD~209458b would require at least $\sim$30
radial velocity measurements with uncertainties of $\sim 2~\ms$
(although, of course, HD~209458 itself has proven to have a much
smaller value of $\lambda$).

For $I_S=60^\circ$ (middle column of Fig.~\ref{fig:drerrs}) the most
favorable value is $\lambda \approx 70^\circ$.  In this case, we find
$\sigma_\lambda=11Q_R^{-1}$, $\sigma_{V_{eq}}/V_{eq}=28Q_R^{-1}$,
$\sigma_\alpha=31Q_R^{-1}$, and $\sigma_{I_S}=42 Q_R^{-1}$. The
variances of the parameters are highly correlated.  In particular,
while the uncertainties in $V_{eq}$ and $I_S$ are individually high,
they are correlated such that the uncertainty in $V_{eq}\sin I_S$ is
considerably smaller.  Measuring all of the parameters to better than
$20\%$ will require $Q_R\ga 200$.

We also computed the uncertainties for the same geometry as above, but
for different values of $I_S$ and the impact parameter $b$. The right
column in Figure \ref{fig:drerrs} shows the results for $b=0.5$ and
$I_S=30^\circ$. Generally we find that the uncertainties are
relatively large for $I_S\la 30^\circ$, primarily because the
amplitude of the RM effect is decreased by the sky projection.
Grazing transits ($b\ga 0.9$) and central transits ($b\approx 0$)
provide poorer constraints, but for $0.2 \la b \la 0.7$, the
uncertainties are relatively insensitive to $b$.

We conclude that, for favorable geometries ($30^\circ \la \lambda \la
70^\circ$ and $0.2 \la b \la 0.7$), it may be possible to detect
differential rotation and to constrain $\alpha$ and $I_S$ to within
$\sim$20\%, using very high signal-to-noise ratio observations of the
RM effect ($Q_R \ga 10^2$).

On the other hand, for lower signal-to-noise ratio observations, we
have shown that the covariances with the differential rotation
parameters (particularly $\alpha$) will degrade the measurement of
$\lambda$. It is is therefore of interest to ask to what degree a
modest {\it a priori} constraint on $\alpha$ will improve the expected
uncertainties on $\lambda$. Because such constraints are not easily
incorporated into the Fisher matrix formalism, we instead determine
the uncertainties using Monte Carlo simulations, as described in
\S\ref{sec:character}.  We first compute the expected uncertainties on
the parameters with no constraints, and so verify our estimates based
on the Fisher matrix formalism.

Again adopting the parameters of the HD~209458b system ($b=0.52$,
$\gamma=0.121$, $\epsilon=0.6$, $\lambda=-4^\circ\hskip-4.5pt.4$,
$V_{eq}\sin I_S=4.70~\kms$, $N=14$, and $\sigma=4.1~\ms$), we find
$\sigma_\lambda=2^\circ\hskip-4.5pt.7$ for $\alpha=0.2$ (with a very
weak dependence on $I_S$). We then enforce a weak constraint on
$\alpha$ by adding a penalty term to $\chi^2$, of the form $\Delta
\chi^2 = [(\alpha-0.2)/0.4]^2$. We find
$\sigma_\lambda=2^\circ\hskip-4.5pt.0$, also with a weak dependence on
$I_S$.  Thus, this mild {\it a priori} constraint reduces the expected
uncertainty in $\lambda$ from a level that is 60\% larger than the
expectation under the assumption of solid-body rotation, to a level
that is only 20\% larger. At higher signal-to-noise ratios, the
constraint on $\alpha$ has little effect.

\section{Confirming Transiting Planets}\label{sec:confirm}

In this section we move from the regime of high $S/N$ to the regime of
low $S/N$, and investigate the utility of the RM effect in confirming
the occurrence of transits in the most challenging cases. The
photometric detection of transits offers one of the most promising
ways to achieve the appealing goal of detecting terrestrial planets in
the habitable zones of solar-type stars. That said, the challenges of
robust detection of the photometric signal are daunting. A transiting
Earth-sized planet orbiting in the habitable zone of a solar-type star
would reduce the stellar flux by only $\sim$10$^{-4}$, for a duration
of only $\sim$13~hr out of the year. Furthermore, the probability that
a randomly oriented orbital plane happens to be close enough to
edge-on to allow transits is only $\sim$0.5\%. The requirements for
detecting these signals have convinced most researchers that it is
necessary to conduct the search with space telescopes rather than
ground-based telescopes. Several satellite experiments are slated for
launch over the next few years that are designed to survey $10^4-10^5$
main-sequence stars in order detect small transiting planets,
including {\it Corot} \citep{baglin03} and {\it Kepler}
\citep{borucki03}.

\begin{figure}[ht*]
\epsscale{1.0}
\plotone{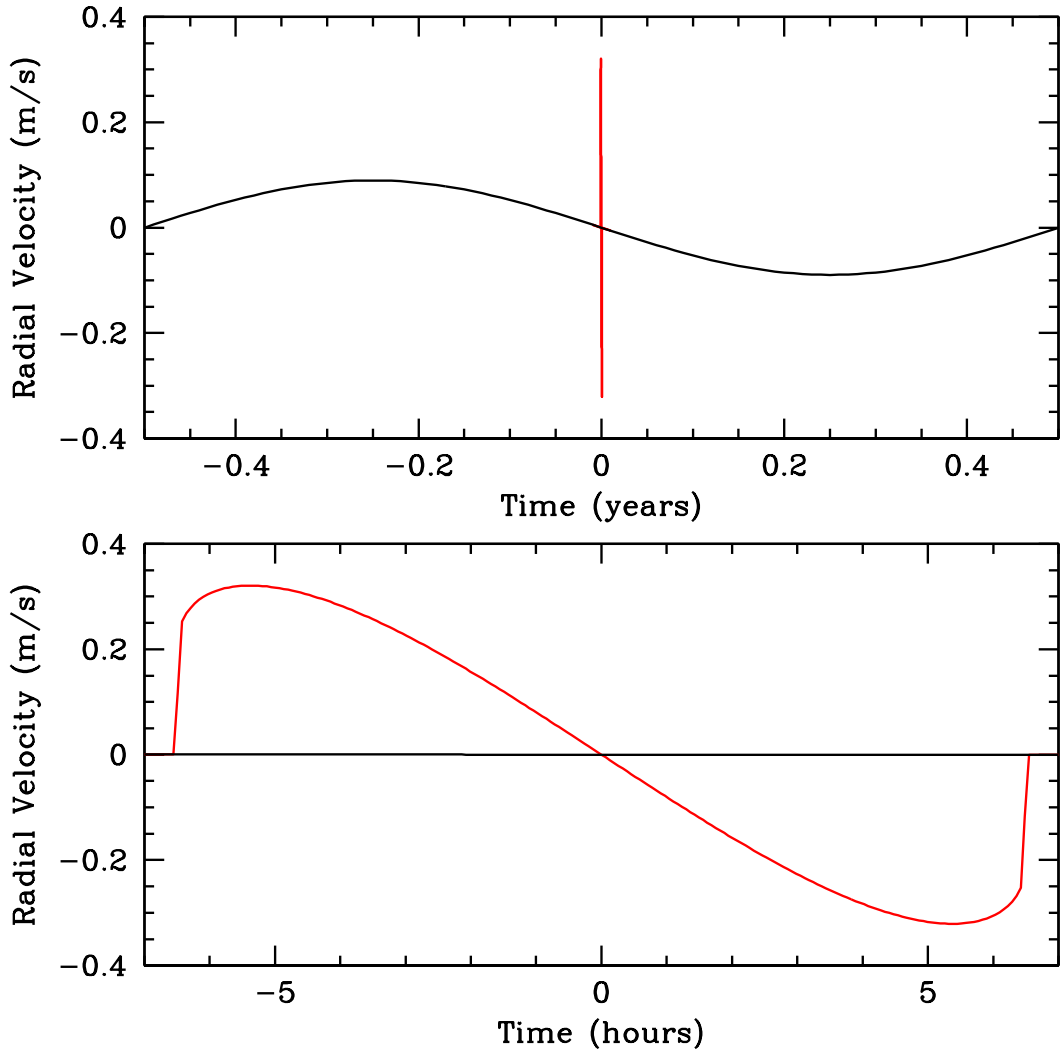}
\caption{ 
Simulated spectroscopic signal of a transiting terrestrial
planet in the habitable zone a solar-type star.  The line-of-sight
component of the stellar orbital velocity is visible as the sinusoid
with a period of one year. The RM effect is visible as the spike at
the origin, which is the time of transit. A close-up of the transit
interval is shown in the lower panel. We have assumed $b=0$,
$\lambda=0$, $\epsilon=0.6$, $M=M_\odot$, $R=R_\odot$, $r=R_\oplus$,
$m=M_\oplus$, $e=0$, $P=1$~yr, and $\vpar = 5~\kms$.
\label{fig:rm-terrestrial}
}\end{figure}

\begin{figure}[ht*]
\epsscale{1.0}
\plotone{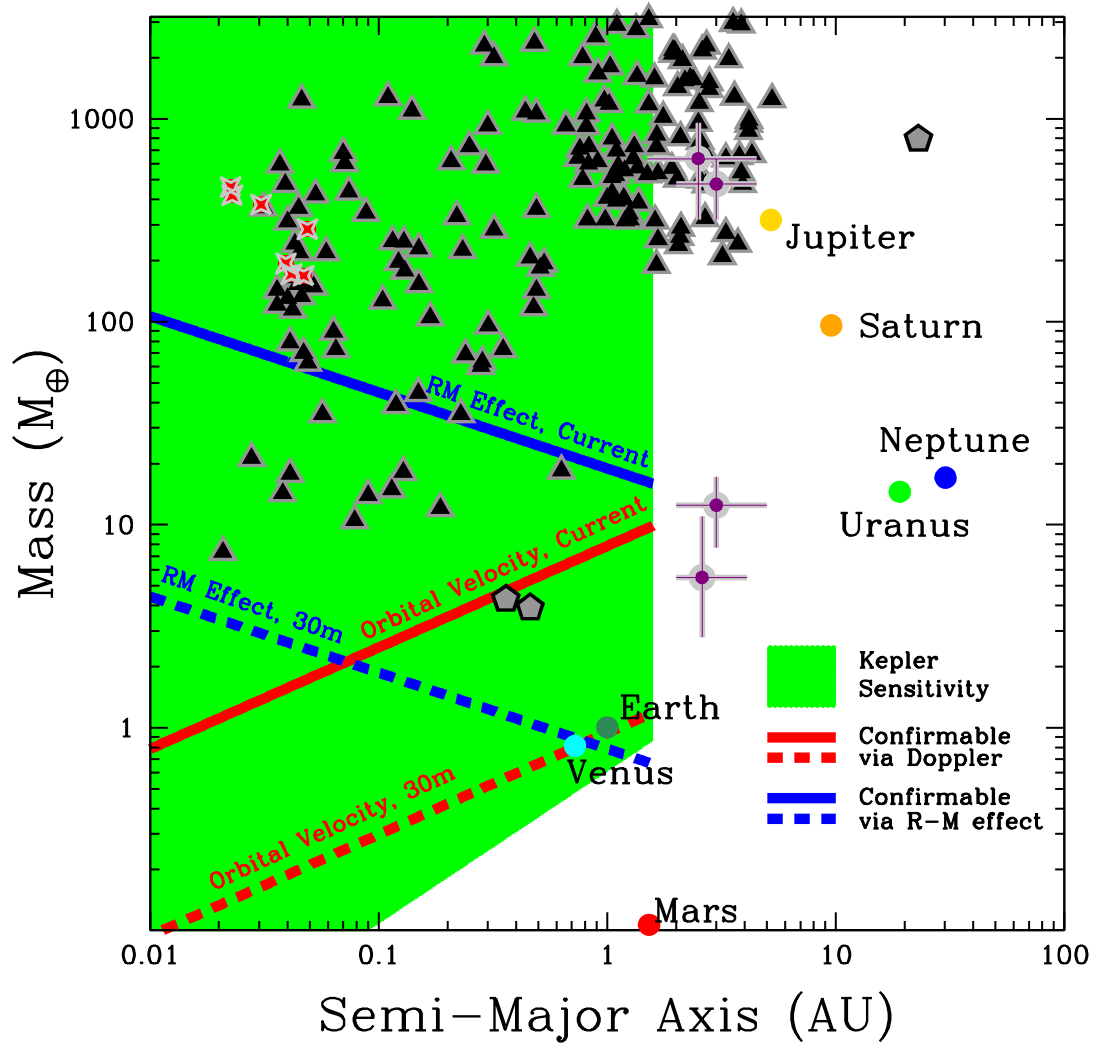}
\caption{{\bf Confirmation of {\it Kepler} planets.}
The symbols show the masses (or minimum masses) of known exoplanets
versus their orbital semimajor axes. The black triangles are planets
detected via spectroscopic orbits, red stars via transits, grey
hexagons via pulsar timing, and purple circles via microlensing. Also
shown are the Solar system planets. The green shaded region is where
{\it Kepler} can detect (with $S/N \ge 8$) at least two transits by a
planet orbiting a star with apparent magnitude $m_V=12$. The lines
show the lower mass limits of planets detectable by Kepler that can be
confirmed (with $S/N \ge 5$) via spectroscopic measurement of the
stellar orbital over the course of a full period (red lines) and via
measurements of the Rossiter-McLaughlin effect throughout a single
transit (blue lines). We show these limits based on the capability of
current instruments such as HARPS (solid lines), and the expected
capability of a future 30m telescope (dashed lines). For this figure,
we have assumed that the host star has the mass and radius of the Sun,
that its projected rotation speed is $5~{\kms}$, and that all planets
have the same mean density. Confirmation via the orbital velocity
gives a lower mass limit in all cases, but it requires $\sim$10 times
more observing time, and greater instrumental stability (the
spectroscopic orbit occurs over a full orbital period, as opposed to a
single transit).
\label{fig:confirm}
 }\end{figure}

As with many such experiments, the most exciting discoveries from
these surveys are likely to be those that are detected with the lowest
signal-to-noise ratio ($S/N$).  Furthermore, the scaling of the number
of detections as a function of the limiting $(S/N)_{\rm min}$ is
generally quite steep.  For example, \citet{gould04} demonstrated that
the number of expected detections for {\it Kepler} scales as
$(S/N)^{-2.4}_{\rm min}$.  This implies that the number of detections
is fairly sensitive to unanticipated degradations in $(S/N)_{\rm min}$
(from unexpected noise sources or other problems), and that the
majority of the detections will occur very near the limiting
$(S/N)_{\rm min}$.  Since the limiting value of $(S/N)_{\rm min}$ will
generally be the minimum possible $S/N$ for which detection is
statistically possible, it will be difficult or impossible to use any
further characteristics of the transit light curves themselves to
distinguish true planets from false positives.  If all of the
information is required for mere detection, then there is not enough
information for the accurate determination of multiple parameters.

Therefore, it will be essential to verify the low $S/N$ candidates
with additional observations.  Unfortunately, since the photometric
noise requirements are so stringent, it may prove very difficult to
confirm these candidates with ground-based photometric observations.
Space-based satellites such as {\it HST} or {\it JWST}
\citep{charbonneau04,gould04} may deliver the sensitivity and
precision to confirm these detections, provided they are available.
However, it would certainly be more expedient to have a good
ground-based method to confirm candidates.

The most desirable type of confirmation would be the spectroscopic
detection of the stellar orbital velocity.  This could be done from
the ground, and the Doppler signal would also provide an estimate of
the mass of the planet, and hence its mean density (when combined with
the radius measurement from the photometric transit). The difficulty
with the detection of the spectroscopic orbit is that the stellar
reflex velocity would be quite small: the aforementioned Earth-like
planet in the habitable zone of a solar-type star produces a
$\sim$9~cm~s$^{-1}$ wobble. Furthermore, the parent stars will be
relatively faint (with an apparent magnitude $m_V \sim 12$ for {\it
  Kepler}), owing to the narrow-field, magnitude-limited design of the
survey observations.

Can current setups detect the spectroscopic orbits of the systems with
habitable planets that will be detected by {\it Kepler}?  Before
answering this question, we first estimate the parameters of the
planets that can be detected with {\it Kepler}.  The signal-to-noise
ratio of a transiting planet with a semimajor axis $a$ and radius $r$
orbiting a star with a radius of $R$ is \beq
\left(\frac{S}{N}\right)_T \simeq (\Gamma T)^{-1/2} \left(\frac{R}{\pi
    a}\right)^{1/2} \left(1-b^2\right)^{1/4}
\left(\frac{r}{R}\right)^2,
\label{eqn:sntransit}
\eeq
where $\Gamma T$ is the total number of photons collected from a star
of the appropriate apparent magnitude during the mission lifetime $T$.
For the parameters of the {\it Kepler} mission, we find
\beq
\left(\frac{S}{N}\right)_T \sim
11
\left(\frac{a}{{\rm AU}}\right)^{-1/2}
\left(\frac{m}{M_\oplus}\right)^{2/3}
\left(\frac{1-b^2}{0.75}\right)^{1/4}
10^{-0.2(m_V-12)},
\label{eqn:snkepler}
\eeq where we have assumed that $R=R_\odot$ and that the planet has
the same density as Earth: $r=R_\oplus(m/M_\oplus)^{1/3}$.

We now estimate the signal-to-noise ratio with which follow-up 
radial velocity measurements can confirm the {\it Kepler} detections.
Assuming a circular orbit, and $N_O$ radial
velocity measurements are made with an uncertainty of $\sigma$
evenly sampled in orbital phase, then the total $S/N$ of the orbital velocity signal is
\begin{equation}
\left(\frac{S}{N}\right)_O = \frac{Q_O}{\sqrt{2}},
\label{eqn:snd}
\end{equation}
where we have defined
\begin{equation}
Q_O  \equiv \sqrt{N_O}\frac{K_O}{\sigma}.
\label{eqn:QD}
\end{equation}

The HARPS spectrograph \citep{pepe02} mounted on the 3.6m ESO
telescope is representative of the state of the art in precision
radial-velocity measurements. \citet{lovis05} discuss observations of
three dwarf stars with the HARPS setup. They take inventory of the
contributions to the total error in HARPS measurements, including the
photon noise, wavelength calibration, guiding errors, and stellar
jitter (which in turn includes both stellar oscillations and
granulation noise). They state that the uncertainties due to
wavelength calibration and guiding errors can probably be improved,
and that it may be possible to average down the noise due to stellar
jitter by using sufficiently long exposures (\citealt{lovis05}, but
see \citealt{bouchy05b}). We will therefore focus on photon noise as
the ultimate limiting noise source. Scaling from the results of
\citet{lovis05}, we can expect HARPS to achieve $\sigma \sim 1~\ms$
precision on a star with apparent magnitude $m_V=12$ in a single
$60~{\rm min}$ exposure, depending on the spectral type of the host
star.  Assuming that $N_O$ radial velocity measurements are made over
the course of the year-long period of a terrestrial planet, this gives
\beq \left(\frac{S}{N}\right)_O \sim 0.6
\left(\frac{N_O}{100}\right)^{1/2} \left(\frac{a}{{\rm
      AU}}\right)^{-1/2} \left(\frac{m}{M_\oplus}\right)
10^{-0.2(m_V-12)} \left(\frac{D}{3.6{\rm m}}\right),
\label{eqn:snrvscale}
\eeq
where we have assumed a primary mass of $M=M_\odot$.
Therefore, confirmation of habitable Earth-mass planets
detected by {\it Kepler} will be difficult with current setups, even
if stabilities of $\sim 0.1~\ms$ over a time scale of $\sim$1~yr can
be achieved.

The requirements for the detection of the RM effect of a transiting
habitable planet are generally less severe, both because the amplitude
of the RM effect is larger than the orbital signal, as discussed in
\S\ref{sec:rmeffect}, and because the time scale over which stability
must be maintained is much shorter. Both of these points are
illustrated in Fig.~\ref{fig:rm-terrestrial}. The $S/N$ of the RM
effect is approximately
\begin{equation}
\left(\frac{S}{N}\right)_R = Q_R \left[\frac{1}{3}(1-4b^2)\cos^2{\lambda}+b^2\right]^{1/2}.
\label{eqn:snrm}
\end{equation}
Note that for $\lambda=0$, $(S/N)_R=Q_R\sqrt{(1-b^2)/3}$, whereas for
$\lambda=90^\circ$, $(S/N)_R = Q_R b$.  Assuming $\sigma \sim
1~\ms$ as above, $M=M_\odot$, $R=R_\odot$, $b=0.5$, 
and continuous measurements during a single transit (i.e.\ $N_R=11.23~{\rm
  hr}/1~{\rm hr}$), we find 
\beq
\left(\frac{S}{N}\right)_R \sim 0.7   
\left(\frac{a}{{\rm AU}}\right)^{1/4}
\left(\frac{m}{M_\oplus}\right)^{2/3}
\left(\frac{\vpar}{5~\kms}\right)
10^{-0.2(m_V-12)}
\left(\frac{D}{3.6{\rm m}}\right),
\label{eqn:snrrmscale}
\eeq Thus the $S/N$ with which the RM effect is measured with only
$N_R=11$ points during transit is roughly equivalent to the $S/N$ with
which the orbital Doppler shift is measured with $N_O=100$ points.
Furthermore, detection of the RM effect only requires stability at the
$\sim$$0.4~\ms$ level for $\sim$11~hours, as opposed to a year.

A downside to RM confirmation is that it must take place on the
particular nights when a transit occurs, making the effort especially
vulnerable to the vagaries of the weather.  And of course, since the
transit only represents a small fraction of the planet's orbit, it is
possible to acquire many more data points on the orbital velocity
curve than on the RM curve; it is possible to gain in the $S/N$ by a
factor of $\sqrt{\pi a /R}$, where $a$ is the planet's semimajor axis.
However, in practice it will be difficult to fully realize this
additional factor, because it is not possible to acquire data
continuously (due to telescope availability, weather, seasonal
observability, and so forth), and because the stability requirements
are harder to achieve for measurements over this longer time baseline.

Nevertheless, the fact that the expected $S/N$ is substantially less
than unity even for continuous observations implies that it will only
be possible to verify the most favorable candidates with current
setups.  Assuming that we can simply scale the photon noise
uncertainties obtained by HARPS to larger apertures, we find that
apertures of $\sim 20-40~{\rm m}$ will be required to confirm [with
$(S/N)_R = 5$] the habitable transiting planets orbiting $m_V=12$
primaries detected by {\it Kepler}.

Figure \ref{fig:confirm} summarizes the prospects for the detection of
planets via {\it Kepler} and the prospects for the confirmation of
those planets via spectroscopic detection of either the orbital
velocity or the RM effect. We show the region of parameter space in
the $m-a$ plane in which {\it Kepler} can detect transiting planets
orbiting stars with apparent magnitude $m_V=12$, assuming at least two
transits are required for detection, and $(S/N)_T \ge 8$.  By design,
{\it Kepler} is (just) sensitive to Earth-mass planets in the
habitable zone. We also show the region of parameter space in which
planets can be confirmed (with $S/N \ge 5$) by detecting either the
orbital velocity or the RM effect using current facilities, and using
the expected capabilities of future 30m telescopes.

\section{Summary and Discussion}\label{sec:discussion}

The transit of an exoplanet is a rare and information-rich event that
should be exploited in every possible way. In this paper we have
examined one aspect of transit physics, the Rossiter-McLaughlin
effect, whose origin is the rotation of the parent star. Our goal has
been to assess the ability of near-term observations to exploit the
information in the RM signal.

We have calculated the achievable accuracy with which one can measure
the key parameter $\lambda$, with which one can assess the degree of
alignment between the stellar spin axis and the planetary orbit, and
have assessed the prospects for the currently known sample of
transiting exoplanets. One might ask further: how accurately must
$\lambda$ be known in order to enable interesting theoretical
advances, such as the discrimination among different planetary
migration theories? In other words, how good is good enough? This
question does not seem to have been addressed in the literature on
planet formation theory, and is an appealing topic for future
research. As suggested in \S~3, for a system with a very hot Jupiter,
even a crude measurement with $\sigma_\lambda=30\arcdeg$ would be
of interest. As a more general benchmark, one would like to achieve at
least enough accuracy to tell whether a given system is similar to the
Solar system, or not. To date, the only system that has been observed
with enough precision for this task is HD~209458, and the result is
that it is Solar-like with a small but significant misalignment of
$\lambda\approx -4\arcdeg$ (Winn et al.~2005). A system could be
different either by having a larger misalignment ($\lambda \gsim
10\arcdeg$), for which an accuracy of a few degrees or more would
suffice, or by having a more perfect alignment, for which an accuracy
of $\sim 0.1\arcdeg$ would be needed.

We have also explained and quantified the potentially important role
of the RM effect in the confirmation of transiting planet candidates.
It is a difficult task, but in many cases it will be easier than the
task of detecting the orbital velocity of the parent star. The
comparative advantages of RM confirmation are that the velocity
amplitude is generally larger, and the {\it acceleration} is certainly
larger. The full range of RM variations occur during a single transit
(typically a single night or two) whereas the orbital velocity
variations occur over the full orbital period (a year, for the
interesting case of an Earth analog). A disadvantage of RM
confirmation is that one does not learn the planetary mass; it is not
a dynamical measurement, {\it per se}, but rather an alternate method
of verifying that a portion of the stellar surface is periodically
occulted. Furthermore, RM observations must take place during
transits, whereas the spectroscopic orbital data can be obtained on
any other nights. In addition, the ease or difficulty of the RM method
depends on the rotation rate of the star, which will vary from system
to system.

Another interesting question for future research is whether or not the
RM effect could ever profitably become the basis of a planet {\it
search} technique, rather than simply a follow-up technique. Ohta et
al.~(2005) suggested that transiting planets might announce themselves
through large outliers in data bases of radial velocity
measurements. The relevant efficiency factors, selection effects, and
detection algorithms for such an undertaking have yet to be worked
out.  Even less certain are the prospects for an RM-based transit
survey, in which stars are spectroscopically monitored specifically
for transits. The obvious drawback is that typical high-resolution
spectrographs examine only one star at a time, whereas photometric
transit surveys observe thousands or even millions of stars at a
time. There are plans for multiplexed Doppler spectrographs in the
near future (Ge et al.~2004), but the multiplexing factor is only
$\sim$50.

\begin{figure}[ht*]
\epsscale{1.0}
\plotone{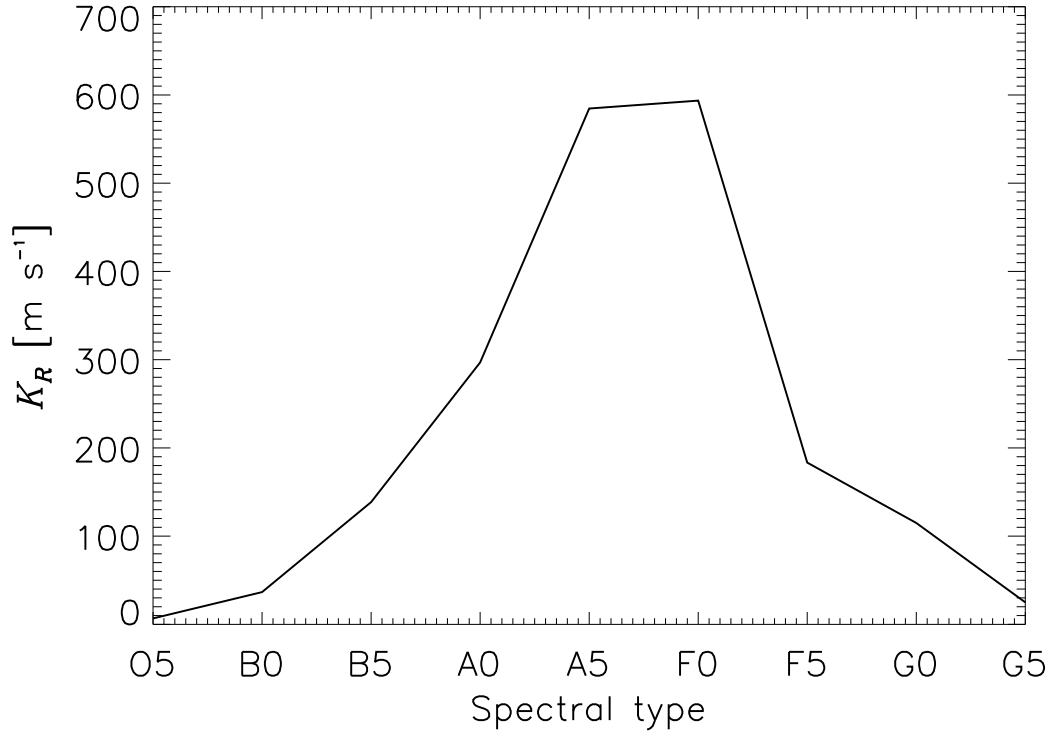}
\caption{
Rough estimate of
the amplitude of the RM effect of a transiting Jovian
planet as a function of the spectral type of the parent star.
The effect reaches a maximum for late A stars, which is a compromise
between the faster rotation rates and the larger radii of hot stars.
\label{fig:kr-vs-spt}
}
\end{figure}

On the other hand, the RM effect is perhaps uniquely sensitive to
small planets around hot and rapidly rotating stars. The amplitude of
the RM effect depends on the stellar radius and projected rotation
speed, according to Eq.~(\ref{eqn:krm}), but these factors are not
completely independent. Among main-sequence stars, both the stellar
radius and the typical rotation speed increase with stellar mass (or
temperature), with an especially dramatic rise near the F5
boundary. The typical rotation speed of G0 stars is
$\sim$10~km~s$^{-1}$, and for F0 stars it is $\sim$100~km~s$^{-1}$
(see, e.g., Tassoul~2000). This is related to the strong variation in
the depth of the convective zone across this same boundary.  We have
attempted to illustrate the net effect of these astrophysical
correlations on the strength of the RM signal, in the following
manner. We considered main-sequence stars ranging in spectral type
from O5 to G0, and imagined that a Jupiter-sized planet makes an
equatorial transit of each star.\footnote{Later-type stars are not
amenable to even this crude analysis, as there is a very wide
dispersion in velocity (and achievable velocity accuracy) that is
linked to age and stellar activity through the relations of
Skumanich~(1972) and Noyes et al.~(1984).} We calculated $K_R$ using
estimates of the stellar radius from Cox et al.~(2000), and estimates
of typical rotation speeds from Tassoul~(2000). The results are shown
in Fig.~\ref{fig:kr-vs-spt}. The maximum signal is achieved for late A
stars.  Of course, the achievable velocity accuracy is also a function
of stellar type, as it depends on the number and width of absorption
lines available for velocity determination. We have not attempted to
quantify the velocity accuracy because in this regime it would
certainly be advantageous to abandon the description of the RM effect
as a net Doppler shift and model the distortion directly, as is done
in Doppler-imaging algorithms.

As with so many other areas of stellar and planetary physics, interest
in the Rossiter-McLaughlin effect has been re-invigorated by the
discovery of exoplanets. It has been nearly 100 years since the first
observations of this effect, and yet we believe that there are still
important applications waiting to be developed.

\acknowledgments We are grateful to Thomas Beatty, Tim Brown, Bill
Herbst, Matt Holman, Ed Turner, and Yasushi Suto for helpful
discussions. B.S.G.\ was supported by a Menzel Fellowship from the
Harvard College Observatory.

\end{document}